\documentclass[useAMS,usenatbib,usegraphicx]{mn2e}
\usepackage{times}
\usepackage{aas_macros}
\usepackage{rotating}
\usepackage{longtable}
\usepackage{times} 

\def\lsim{~\raise0.3ex\hbox{$<$}\kern-0.75em{\lower0.65ex\hbox{$\sim$}}~}
\def\gsim{~\raise0.3ex\hbox{$>$}\kern-0.75em{\lower0.65ex\hbox{$\sim$}}~}

\def\gs{\mathrel{\raise0.35ex\hbox{$\scriptstyle >$}\kern-0.6em \lower0.40ex\hbox{{$\scriptstyle \sim$}}}}
\def\ls{\mathrel{\raise0.35ex\hbox{$\scriptstyle <$}\kern-0.6em \lower0.40ex\hbox{{$\scriptstyle \sim$}}}}

\newcommand{\Msolar}{\mbox{$\rm M_{\odot}\,$}}
\newcommand{\Lsolar}{\mbox{$\rm L_{\odot}\,$}}

\newcommand{\fdr}{S_{\rm 850\mu m}/S_{\rm 1200\mu m}}

\begin{document}

\title[A MAMBO survey of the GOODS-N field]
      {A 1200-$\mu$m MAMBO survey of the GOODS-N field:
       a significant population of submillimetre drop-out galaxies}

\author[Greve et al.]{
\parbox[t]{\textwidth}{
\vspace{-1.0cm}
Thomas R.\ Greve,$^{\! 1,2}$ Alexandra Pope,$^{\! 3,4}$ Douglas Scott,$^{\! 3}$
Rob.\,J.\ Ivison,$^{\! 5,6}$ Colin Borys,$^{2}$ Christopher J.\ Conselice$^{7}$ and
Frank Bertoldi$^{8}$
}
\vspace*{6pt}\\
$^1$Max-Planck Institute f\"ur Astronomie, K\"onigstuhl 17, Heidelberg 69117, Germany\\
$^2$California Institute of Technology, Pasadena, CA\,91125, USA\\
$^3$Department of Physics \& Astronomy, University of British Columbia, Vancouver, BC V6T1Z1, Canada\\
$^4$Spitzer Fellow; National Optical Astronomy Observatory, 950 North Cherry Avenue, Tucson, AZ 85719, USA\\
$^5$UK Astronomy Technology Centre, Royal Observatory, Blackford Hill, Edinburgh EH9 3HJ\\
$^6$Institute for Astronomy, University of Edinburgh, Blackford Hill, Edinburgh EH9 3HJ\\
$^7$University of Nottingham, School of Physics \& Astronomy, Nottingham NG7 2RD\\
$^8$Argelander Institute for Astronomy, University of Bonn, Auf dem H\"ugel 71, 53121 Bonn, Germany
\vspace*{-0.5cm}}

\date{\fbox{\sc Draft dated: \today\ }}
\date{Accepted ... ; Received ... ; in original form ...}

\pagerange{000--000}

\maketitle

\begin{abstract}
We present a 1200-$\mu$m image of the Great Observatories Origin Deep
Survey North (GOODS-N) field, obtained with the Max Planck Millimeter
Bolometer array (MAMBO) on the IRAM 30-m telescope. The survey covers
a contiguous area of 287\,arcmin$^2$ to a near-uniform noise level of
$\sim$0.7\,mJy\,beam$^{-1}$. After Bayesian flux deboosting, a total
of 30 sources are recovered ($\ge$3.5-$\sigma$). An optimal
combination of our 1200-$\mu$m data and an existing 850-$\mu$m image
from the Submillimetre Common-User Bolometer Array (SCUBA) yielded 33
sources ($\ge$4-$\sigma$).  We combine our GOODS-N sample with those
obtained in the Lockman Hole and ELAIS\,N2 fields (Scott et al.\ 2002;
Greve et al.\ 2004) in order to explore the degree of overlap between
1200-$\mu$m- and 850-$\mu$m-selected galaxies (hereafter SMGs),
finding no significant difference between their $\fdr$
distributions. However, a noise-weighted stacking analysis yields a
significant detection of the 1200-$\mu$m-blank SCUBA sources, $\fdr =
3.8\pm 0.4$, whereas no significant 850-$\mu$m signal is found
for the 850-$\mu$m-blank MAMBO sources ($\fdr = 0.7\pm 0.3$).
The hypothesis that the $\fdr$ distribution of SCUBA sources is also
representative of the MAMBO population is rejected at the
$\sim$4-$\sigma$ level, via Monte Carlo
simulations. Therefore, although the populations overlap, galaxies selected at 850 and 1200\,$\mu$m are
different, and there is compelling evidence for a significant
1200-$\mu$m-detected population which is not recovered at 850\,$\mu$m.
These are submm drop-outs (SDOs), with $\fdr = 0.7-1.7$, requiring very cold
dust or unusual spectral energy distributions ($T_{\rm d}\simeq 10$\,{\sc
k}; $\beta \simeq 1$), unless SDOs reside beyond the redshift range observed
for radio-identified SMGs, i.e.\ at $z > 4$.
\end{abstract}


\section{Introduction}

A decade ago, surveys undertaken at submillimetre (submm; 850-$\mu$m)
wavelengths with SCUBA (Holland et al.\ 1999) transformed the accepted
understanding of galaxy formation and evolution by revealing an
unexpected yet significant population of dusty, far-infrared-luminous
starburst galaxies (Smail, Ivison \& Blain 1997; Hughes et al.\ 1998;
Barger et al.\ 1998).  Since then, great strides have been made in
understanding the nature of submm-selected galaxies -- see Blain et
al.\ 2002 for a review -- largely facilitated by accurately
pinpointing SMGs in deep, high-resolution radio maps (e.g.\ Ivison et
al.\ 1998a, 2000, 2002; Smail et al.\ 2000) which helps overcome the
coarse resolution of SCUBA ({\sc fwhm} $\simeq$ 15\,arcsec). The
majority (60--80 per cent) of bright ($S_{\rm 850\mu m}\gs 5$\,mJy)
SMGs have been identified in this way and one of the most important
steps forward was the acquisition of about 100 optical/near-infrared
spectra of such radio-identified SMGs. This placed them typically in the
redshift range $z\simeq 1-3$, with a median of 2.3 (Chapman et al.\
2003, 2005).  However, while submm surveys are equally sensitive to
sources of a fixed luminosity in the redshift range $z\simeq 1-8$, due
to the negative $k$-correction offsetting the cosmic dimming (Blain \&
Longair 1993), the detection rate of SMGs in even the deepest radio
surveys drops off rapidly beyond $z\simeq 3.5$. Thus, the
spectroscopic requirement for a robust radio detection biases the
known population to $z \ls 3.5$. This led to the suggestion that the
20--40 per cent of the bright, radio-blank SMGs (1.4\,GHz flux density, $S_{\rm 1.4GHz} \ls
20$\,$\mu$Jy) could be the high-redshift tail of the population ($z >
4$ -- see Carilli \& Yun 1999; Ivison et al.\ 2002; Eales et al.\
2003; Aretxaga et al.\ 2003, 2007), although using photometric redshifts
Pope et al.\ (2006) argued that $<$14 per cent of the bright SMG population is at $z > 4$.

Perhaps the strongest candidates for $z > 4$ SMGs have come from
submm/mm interferometry, with the IRAM Plateau de Bure Interferometer
(PdBI) and the Submillimeter Array (SMA), of a few very bright SMGs
which -- by virtue of their radio, submm/mm and near-infrared
properties -- were deemed to reside at very high redshift (Dannerbauer
et al.\ 2002; Wang et al.\ 2007; Younger et al.\ 2007).

Soon after the first observations with SCUBA, surveys were being
conducted at 1200\,$\mu$m with MAMBO (Bertoldi et al.\ 2000) resulting
in samples of sources that were thought to be identical to SMGs,
except selected at mm wavelengths. Compared to submm surveys, however,
observations at mm wavelengths should be sensitive to far-IR-luminous
galaxies out to higher redshifts, due to the $k$-correction, as well
as to systems with cooler dust temperatures.  As a result, at $z\gs 3$
the 850-/1200-$\mu$m flux ratio becomes a relatively strong function
of redshift for a typical starburst far-infrared/mm spectral energy
distribution (SED), and may therefore (if the SED is known) be used as
a crude redshift estimator or -- if the redshift is known -- as an
indicator of the shape (i.e.\ temperature and spectral index) of the
far-infrared/mm SED.

This fact was exploited by Eales et al.\ (2003; hereafter E03), who obtained
850-$\mu$m photometric measurements of a sample of
1200-$\mu$m-selected sources and found a significant fraction to have
very low 850/1200-$\mu$m flux ratios ($\fdr \ls 2$). They argued that
this was either indicative of SMGs at $z \gg 3$ or -- since a large
fraction of their sample was radio-identified with radio-to-submm flux
ratios consistent with $z \le 3$ -- SMGs having fundamentally
different dust emission properties than local galaxies. The existence
of a significant number of these SDOs, which we define here as sources
with $S_{\rm 850\mu m}/S_{\rm 1200\mu m} \ls 2$, was questioned
(albeit not ruled out) by Greve et al.\ (2004; hereafter G04), who
performed the first unbiased comparison of 850- and
1200-$\mu$m-selected sources, using catalogues extracted from SCUBA
and MAMBO maps.  The degree of overlap between source populations
extracted from submm and mm surveys (and in particular the idea that
mm observations are tracing a significant sub-population of more
distant and/or cooler sources) remains controversial, therefore, and
is further complicated by the difficulty of comparing (sub)mm surveys
with different depths and noise properties. Nonetheless, settling this
issue is important; regardless of the outcome, it tells us about the
population of dusty galaxies at the highest redshifts, with
significant implications for our understanding of galaxy formation and
evolution.

In this paper we present an unbiased 1200-$\mu$m MAMBO survey of the
northern field of the Great Observatories Origins Deep Survey (GOODS-N
-- Giavalisco et al.\ 2004). As well as presenting the 1200-$\mu$m map
and source catalogue, we also perform a rigorous comparison between
our data and the existing 850-$\mu$m SCUBA map of this region (Borys
et al.\ 2003; Pope et al.\ 2005), in order to address the issues
outlined above.  In a follow-up paper, we shall exploit other
high-quality data in the GOODS-N field to explore the multi-wavelength
properties of 1200-$\mu$m selected sources.

Throughout this paper we have adopted a flat cosmology with
$\Omega_{\rm M} = 0.27$, $\Omega_{\Lambda}=0.73$, and
$H_0=71\,\mbox{km}\,\mbox{s}^{-1}\,\mbox{Mpc}^{-1}$ (Spergel et al.\
2003).

\section{Observations and data reduction}
\label{section:observations-and-data-reduction}

%
%
\begin{figure*}
\begin{center}
\includegraphics[width=0.65\hsize,angle=-90]{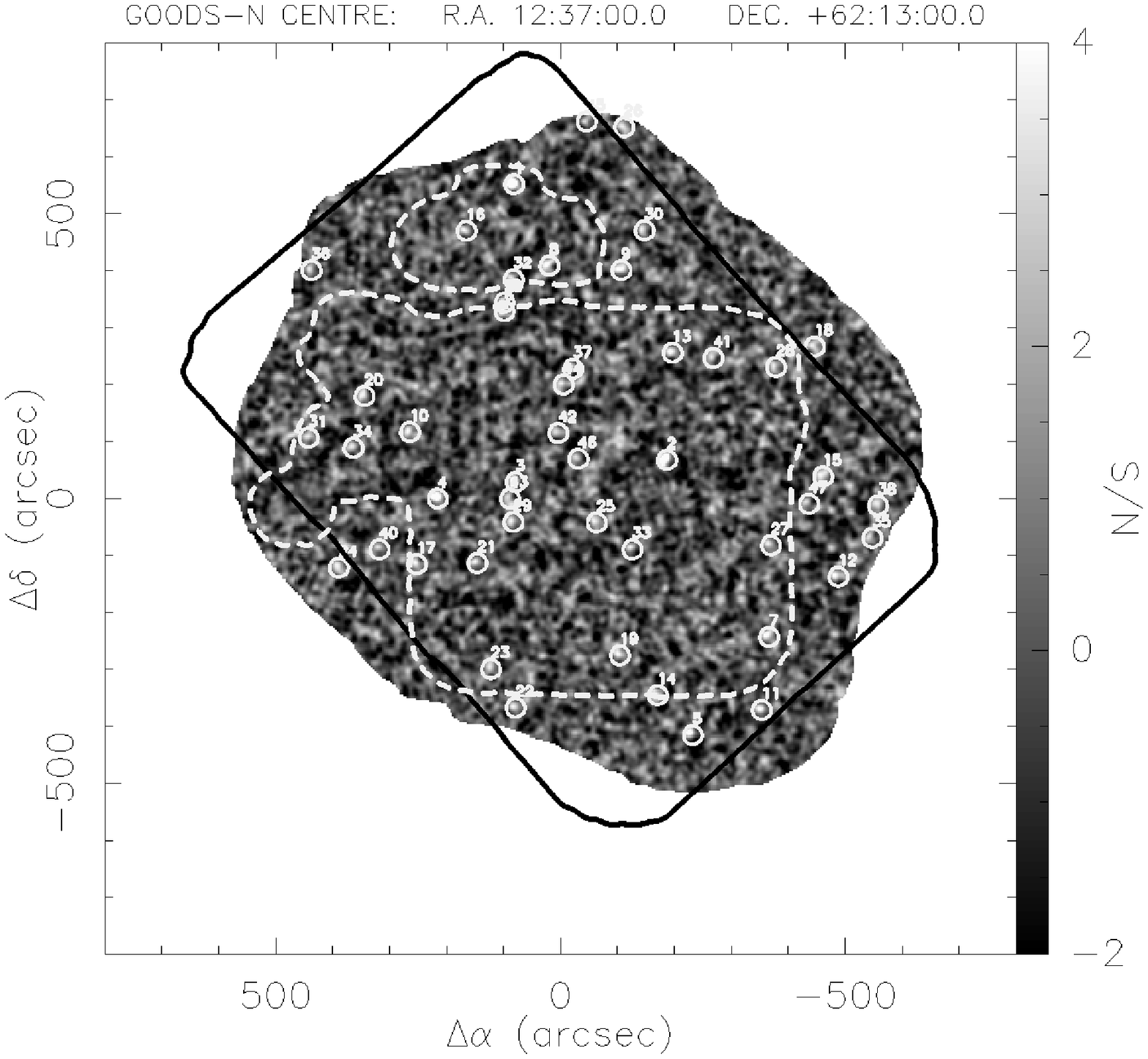}
\includegraphics[width=0.65\hsize,angle=-90]{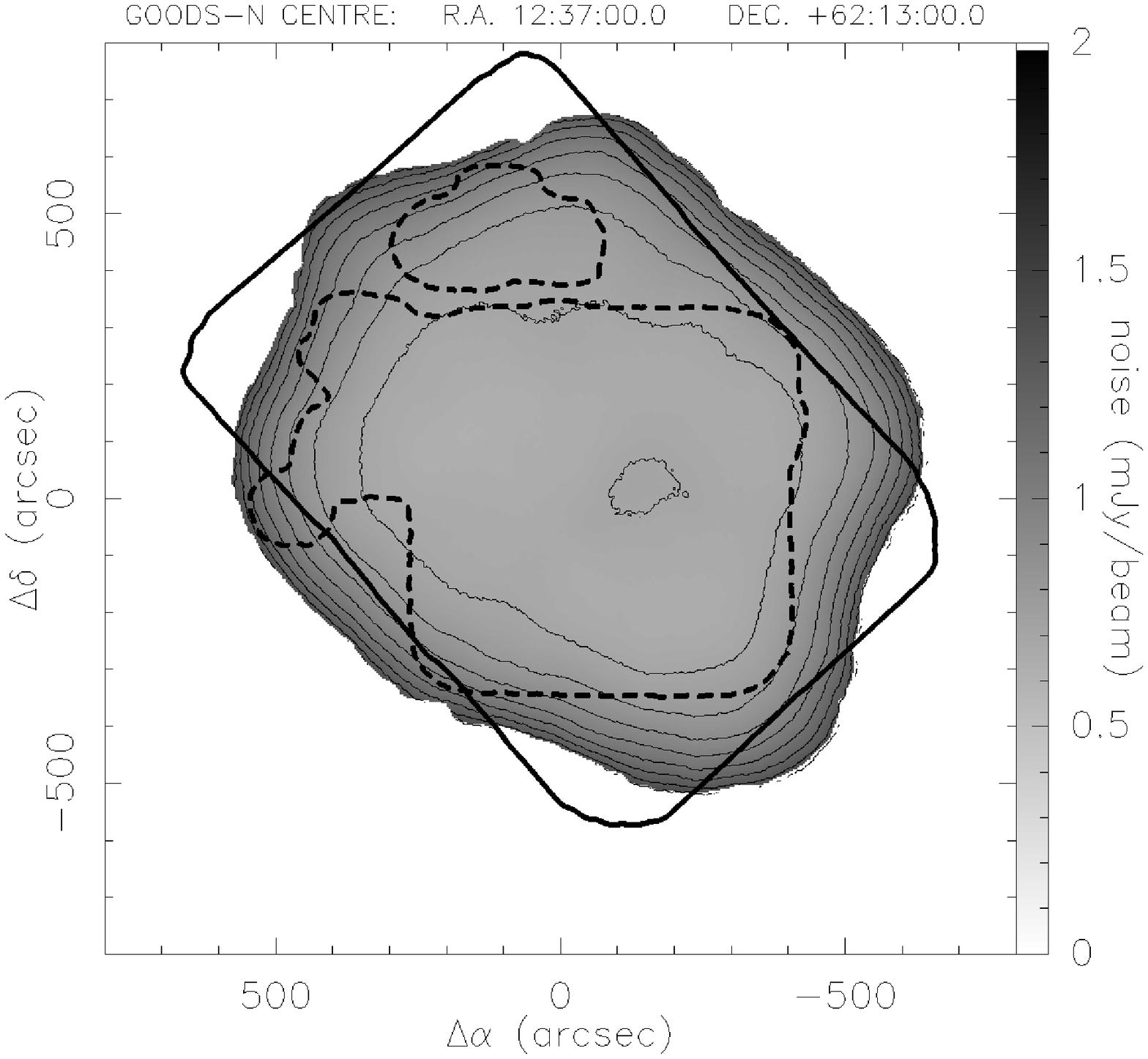}
\caption[]{{\bf Top:} The 1200-$\mu$m MAMBO S/N map of the GOODS-N
region. Sources extracted at $\ge$3.5\,$\sigma$ (prior to deboosting,
see \S \ref{subsection:flux-boosting}) are indicated as white
circles. {\bf Bottom:} The corresponding r.m.s.\ noise map. The thin
black contours correspond to $\sigma = 0.7,...\,,1.2$\,mJy\,beam$^{-1}$.
The thick black outline shows the full extent of the GOODS-N field as
observed with {\it Spitzer}/IRAC and the thick dashed line shows the
extent of the SCUBA map (Borys et al.\ 2003; Pope et al.\ 2005).  }
\label{figure:mambo-maps}
\end{center}
\end{figure*}

A 1200-$\mu$m (250-GHz) map covering the entire GOODS-N region was
obtained using the 117-channel bolometer array, MAMBO (Kreysa et al.\
1998), on the IRAM 30-m telescope on Pico Veleta in Spain. The data
were obtained during the winter semester 2005-06 in good weather
conditions, meaning 250-GHz opacities of $\ls$0.3 and stable
atmospheric conditions (low sky-noise).

In order to map the entire field as uniformly as possible, a number of
regular grid positions (2 arcmin apart) were observed.  Specifically,
each such grid position was observed twice, using a standard
on-the-fly 300\,arcsec $\times$ 320\,arcsec Az--Alt scan-map.  With a
scanning speed of 5\,arcsec\,s$^{-1}$ in Az and an elevation spacing
between each sub-scan of 8\,arcsec, a scan-map takes $\sim$43\,min to
complete, allowing for a 3-s turnaround after each sub-scan.  To avoid
any systematics and to minimise the effects of sky rotation, we tried
to observe each grid position once when the field was rising and once
when it was setting. Different wobbler throws (36--45\,arcsec) and
scan directions (increasing or decreasing elevation) were also
employed for different scan-maps. In total, 53 scan-maps were
obtained.

After each scan-map, a pointing observation on a nearby, bright point
source was carried out and, in order to be extra cautious, a focus
correction was usually applied after each scan (sometimes after two
scans if the focus appeared stable). Since the weather conditions were
generally stable, it was sufficient to perform a skydip every 2\,hr in
order to monitor the 250-GHz atmospheric opacity. Skydips were carried out at
the same azimuth as the preceeding grid-point observation, ensuring
that the opacity was measured for the same part of the sky as the
science observations.

The data were reduced with the {\sc mopsic} software -- an improved
version of the original {\sc mopsi} package (Zylka 1998) -- using
standard pipeline scripts designed to deal with MAMBO data.  This
involved many of the standard steps in reducing bolometer data, i.e.\
de-spiking, flat-fielding, removal of correlated noise across the
array (see G04 for details), and correction for atmospheric opacity by
interpolating between the skydip measurements.  The timestream data
were rebinned onto a 1\,arcsec $\times$ 1\,arcsec grid using a shift-and-add technique,
which results in each positive source being bracketed by two negative
counter-images, each with half of the peak positive intensity, located
one wobbler throw away, i.e.\ the standard `double-difference' or
`triple-beam' pattern.  However, the different wobbler throws and
scanning directions employed for different scans ensure that the
negative images are averaged out, resulting in a very clean image
(although with some residual effects, discussed in \S~\ref{subsection:source-extraction}).

The final 1200-$\mu$m signal-to-noise (hereafter S/N) map and
corresponding noise map are shown in Fig.~\ref{figure:mambo-maps}. The
area covered is 287\,arcmin$^2$.  The map exhibits a fairly uniform
r.m.s.\ noise of $\sim$0.7\,mJy\,beam$^{-1}$ over most of its area
(apart from at the edges where the noise is larger,
$\gs$0.8\,mJy\,beam$^{-1}$).

%
%
\begin{figure*}
\begin{center}
\includegraphics[width=1.0\hsize,angle=0]{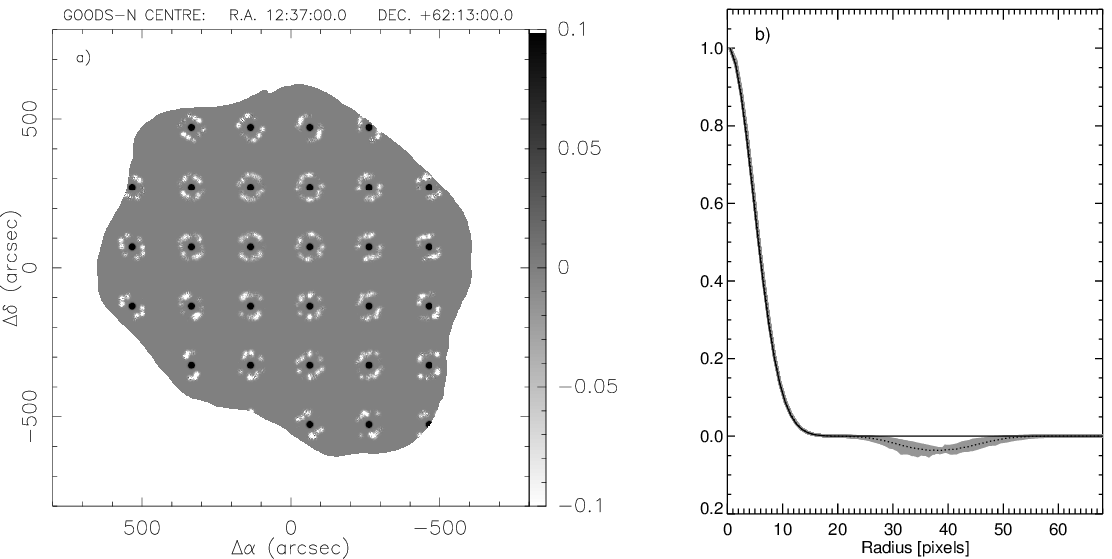}
\caption[]{{\bf a)} Variations in the PSF at a collection of grid
points across the map. The negative parts of the PSF (shown in white)
are caused by the wobbler throw used in the on-the-fly mapping mode on
the IRAM 30-m telescope. Due to sky rotation and variations in the
chop throw (between 36 and 45\,arcsec) the negative residuals are
smeared out in a ring around the main beam. {\bf b)} Azimuthally
averaged PSF profiles. The gray-shaded region shows the radially
averaged profiles of all the PSFs plotted on the left, and the dotted
line is the average of those profiles. The solid line is a best-fit
Gaussian with a fixed {\sc fwhm} of 11.1\,arcsec.  }
\label{figure:psf-map}
\end{center}
\end{figure*}

\section{The GOODS-N 1200-micron source catalogue}
\label{section:source-catalogue}

\subsection{Source extraction}
\label{subsection:source-extraction}

In order to extract sources, a point spread function (PSF) was fitted
to each pixel in the map and a S/N ratio was ascribed, based on the
local noise.  This is similar to the noise-weighted convolution
technique applied to MAMBO maps by G04 and to SCUBA maps by Borys et
al.\ (2003) and Coppin et al.\ (2006).  Unlike SCUBA, where one can
configure the nod/chop such that it remains constant on the sky, MAMBO
experiences sky rotation; as a result, the PSF varies across the map.

We have gone to great lengths to average out the negative chop
patterns (see \S~\ref{section:observations-and-data-reduction}), but
we still expect to see some effects of the chop throw in the map. To
quantify this, we created fake, noiseless time streams of data,
containing a number of point sources at certain positions in the map,
and then subjected these data to exactly the same data-reduction
pipeline as the real data, using the astrometrical information
(coordinates, chop throws etc.) of the real data to create a map which
shows how the PSF varies across the map (Fig.~\ref{figure:psf-map}a).

Ideally, one would want to use the PSF (including its negative parts)
as the filtering kernel in the source-extraction procedure, as has
been done with SCUBA maps (e.g.\ Scott et al.\ 2002, hereafter S02;
Borys et al.\ 2003; Coppin et al.\ 2006).  In our case, that would
entail computing a separate PSF for every pixel in the map, or at
least some sub-region of the map where the PSF is approximately
constant. However, as seen in Fig.~\ref{figure:psf-map}a, the negative
features of the net PSF are fairly low contrast, spread out over a
large area, making it difficult (and also largely unnecessary) to use
them explicitly in the source-extraction procedure.

This also suggests that the optimal method of combining the data,
using an explicit PSF for each individual time sample (see Borys et
al.\ 2003), would produce only marginally better results, at the
expense of enormously increased complexity. Indeed, we used a few
sources to show that the improvement was modest.  We therefore decided
to use a simple Gaussian ({\sc fwhm}, 11.1\,arcsec) as an
approximation to the PSF. Although the 30-m telescope's beam is
formally 10.7\,arcsec {\sc fwhm} at 1200\,$\mu$m, we have adopted a
slighter broader PSF in order to account for the typical pointing
error ($\sim$3\,arcsec r.m.s.).  That this is, in fact, a good
approximation is demonstrated in Fig.~\ref{figure:psf-map}b where we
have plotted the mean azimuthally averaged PSF profiles, along with a
Gaussian profile with {\sc fwhm} = 11.1\,arcsec. While half of the
signal is contained in the negative residuals, it is seen that the
depth of the negative holes reach only about 5 per cent of the primary
PSF height. This procedure yielded 20 sources with S/N $\ge$ 4 and a
further 27 sources with $3.5 \le {\rm S/N} < 4$ (see
Table~\ref{table:mambo-source-list}).

\subsection{Monte Carlo simulations}
\label{subsection:monte-carlo-simulations}

\subsubsection{Spurious sources}
\label{subsubsection:spurious-sources}

\begin{figure}
\begin{center}
\includegraphics[width=1.0\hsize,angle=0]{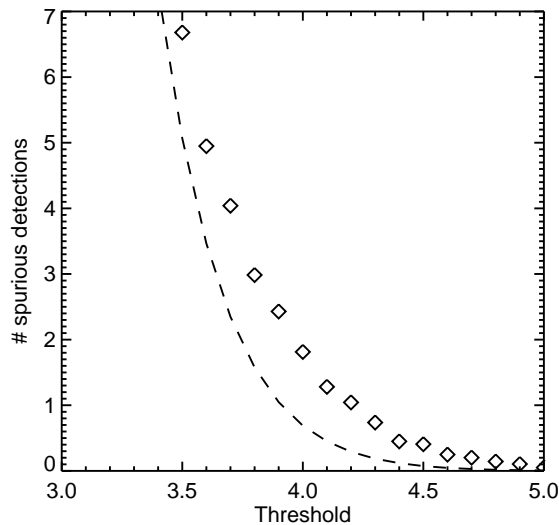}
\caption[]{Diamonds show the average number of (positive) spurious
detections as a function of S/N ratio, based on simulated maps with
realistic noise properties.  The dashed line shows the expected number
of spurious sources, assuming purely Gaussian noise and a Gaussian
beam with a {\sc fwhm} of 11.1\,arcsec.  }
\label{figure:spurious-detections}
\end{center}
\end{figure}

In order to assess the degree to which the extracted source catalogue
is contaminated by spurious noise peaks, we randomised the bolometer
positions on the array in order to create a scrambled map with no
sources, with noise properties similar to that of the real map. In
total, 500 such scrambled maps were produced, each processed by the
data-reduction pipeline in exactly the same way as the real
data. Fig.~\ref{figure:spurious-detections} shows the average number
of spurious detections in the scrambled maps, when subjected to our
source-extraction technique.  The simulations suggest that for
$\ge$4-$\sigma$ sources extracted from the 287\,arcmin$^2$ surveyed,
less than two are expected to be spurious, while we can expect about
seven spurious sources in the 3.5-$\sigma$ catalogue.  This is in
agreement with similar simulations by
G04. Fig.~\ref{figure:spurious-detections} also shows how the expected
number of spurious sources is larger than for purely uniform Gaussian
noise, confirming that it is important to carry out the simulations
with realistic noise.

A further check on the source robustness can be made by extracting
sources from an inverted map, in other words, carrying out the same
extraction procedure for {\it negative} sources, which of course can
only be real when created by freak alignments of sources and chop
throws. In total, we find 33 sources above 3.5\,$\sigma$ in the
inverted map. All but six are associated with the negative beams of
bright positive sources, consistent with expectations.

%
%

\subsubsection{Completeness and positional offsets}

Catalogue completeness varies as a function of the source flux,
becoming smaller at lower flux values. In addition, random noise
fluctuations result in the extracted positions being offset from the
true source positions; again, the effect becomes increasingly severe
at lower fluxes. The effect of the noise on the completeness of the
survey and on the positional offsets was assessed using Monte Carlo
simulations.  Artificial sources with fluxes in the range 1--12\,mJy
(in steps of 0.25\,mJy) were added to the science map, one at a time,
at random positions, and subsequently extracted using exactly the same
technique used in \S~\ref{subsection:source-extraction}.  This
procedure was repeated 50 times for every flux bin, each time at a
random position in the map. Each injected source was a scaled version
of the local PSF and as a result the shape and depth of the negative
lobes around the sources was a function of the map position.  We also
excluded artificial sources which were within half a beam width of any
real source, hence this definition of completeness is intended only to
assess the effects of noise, without including the effects of
confusion (Blain, Ivison \& Smail 1998).

%
%
\begin{figure}
\begin{center}
\includegraphics[width=1.0\hsize,angle=0]{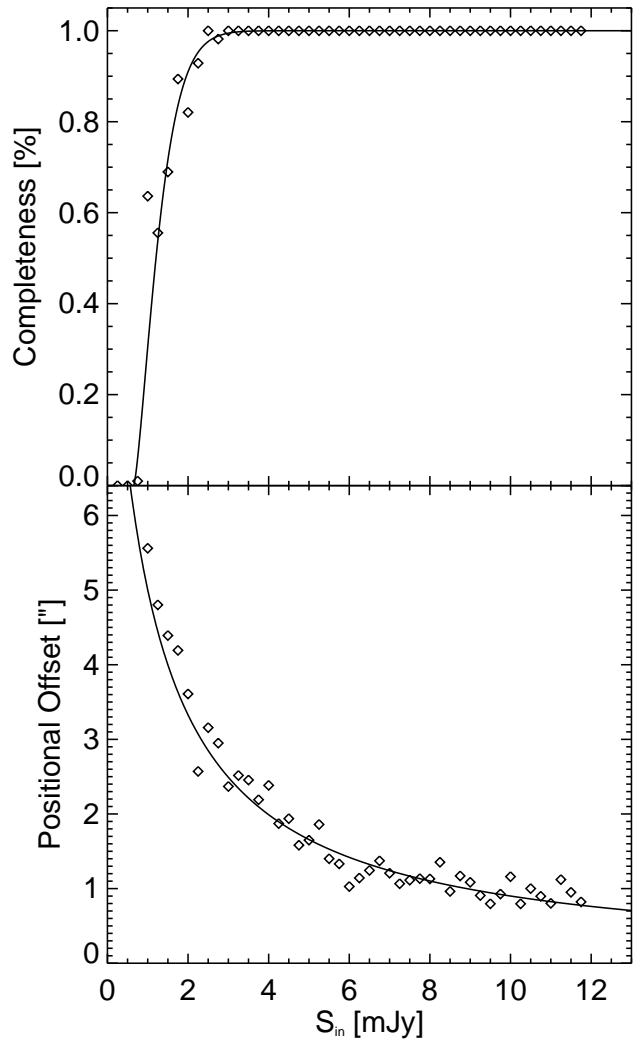}
\caption[]{{\bf Top:} The completeness function of our survey, derived
as the percentage of input sources which are recovered by the source
extraction as a function of the source flux density (but not including
the effects of confusion). The solid curve represents a fit to the
phenomenological expression: $f(S_{\rm in}) = 1- {\rm exp}(A(S_{\rm
in}-B)^C)$. {\bf Bottom:} The positional offsets between the actual
positions of the input sources and their extracted position.  A fit to
the function $f(S_{\rm in}) = 1+A \,{\rm exp}(B\,S_{\rm in}^C)$ is shown by
the solid curve.}
\label{figure:monte_carlo}
\end{center}
\end{figure}

The resulting completeness curve and positional offsets are shown in
Fig.~\ref{figure:monte_carlo}.  The simulations suggest that the
survey is close to complete down to a flux density of about 3\,mJy,
and drops precipitously below 2.5\,mJy. The lower panel of
Fig.~\ref{figure:monte_carlo} shows that the positions of sources with
flux densities larger than 2.5\,mJy are determined to within
3\,arcsec. Of course, in addition to the positional offsets shown in
Fig.~\ref{figure:monte_carlo}, the data will contain offsets due to
pointing errors. The completeness fractions and positional offsets we
obtain here are very similar to those found by G04.

\subsubsection{Flux deboosting}
\label{subsection:flux-boosting}

%
%
\begin{figure*}
\begin{center}
\includegraphics[width=1.0\hsize,angle=0]{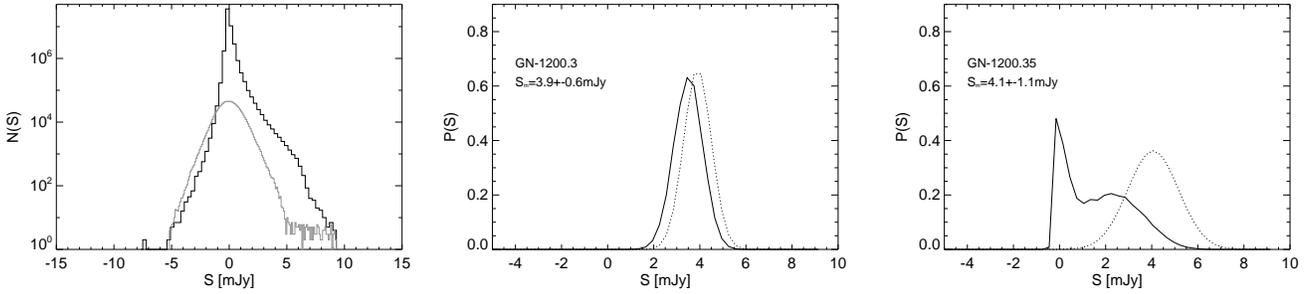}
\caption[]{{\bf Left:} The prior flux density probability
distribution, $N(S_{\rm p})$, of 1,000 simulated maps. Whilst strongly
peaked around zero, the distribution is skewed towards positive fluxes
due to the presence of sources in the maps. The flux density
distribution of the real map is shown in grey. {\bf Middle:} Observed
flux density probability distribution (dotted curve) assuming Gaussian
errors, $P_M(S_{\rm p})$, and the posterior flux density probability
distribution (dark solid curve), $P(S_{\rm p})$, for the source
MM\,J123711$+$621328 (GN\,1200.3), which is well detected.  {\bf
Right:} An example of a source, MM\,J123541$+$621150 (GN\,1200.35),
which is less well detected. In this case the posterior flux density
distribution peaks at zero and the posterior probability for having
zero flux density is larger than 5 per cent.  Hence this source does
not meet our detection criterion, after deboosting.  }
\label{figure:deboost}
\end{center}
\end{figure*}

For any flux-limited survey in the low-S/N regime (as is the case for
all submm/mm surveys undertaken to date), we must worry about the
superposition of sources with positive noise peaks in the map: `flux
boosting' (S02). This effect is particular significant in the submm/mm
waveband where the source number counts have a steep negative slope,
so that the noise excursions above a threshold do not balance
excursions below the threshold.

We chose to follow the Bayesian flux deboosting approach developed by
Coppin et al.\ (2005) to deal with SCUBA data.  The philosophy of this
method is to assume {\em a priori} knowledge of the pixel brightness
distribution, $N(S_{\rm p})$, in the {\it noiseless} map.  If the
noise in the map is Gaussian, the probability of measuring a flux
density $S_{\rm m}\pm \sigma_{\rm m}$ at some pixel in the map when
its true flux is $S_{\rm p}$, is
\begin{equation}
P_M(S_{\rm p}) \propto \exp \left ( {\frac{(S_{\rm m}-S_{\rm
p})^2}{2\sigma_{\rm m}^2}} \right ).
\end{equation}
We can then use Bayes' theorem to derive the probability that the true flux
of the pixel is $S_{\rm p}$:
\begin{equation}
P(S_{\rm p}) = \frac{N(S_{\rm p}) P_M(S_{\rm p})}{P(M)},
\end{equation}
where $P(M)$ is simply the normalisation constant.  The {\em a priori}
flux density distribution, $N(S_{\rm p})$, was generated by making
noiseless maps onto which sources selected from a model of the number
counts were randomly distributed, and then sampled with the adopted
chop pattern.  In our case, the chop pattern varies across the map so
the overall $N(S_{\rm p})$ distribution was derived from noiseless
maps created with exactly the same scanning strategy as the real map.
For the number counts, we adopted a Schechter function of the form
$dN/dS \propto (S/S_0)^{-\alpha} {\rm exp}(-S/S_0)$, where
$S_0=4$\,mJy and $\alpha=2.3$; this provides a good fit to the
1200-$\mu$m number counts by G04.  The $N(S_{\rm p})$ distribution
derived from 1,000 such maps is shown in the left panel of
Fig.~\ref{figure:deboost}.  We checked that the use of moderately
different number count models has little effect on our results; one
has to considerably alter the faint-end slope in order to make
significant changes.  Examples of deboosting are given in
Fig.~\ref{figure:deboost}, which shows the probability distribution of
the measured flux $P_M(S_{\rm p})$ (assuming it is Gaussian) and the
corresponding deboosted flux distribution $P(S_{\rm p})$ for the two
sources, GN\,1200.3 and GN\,1200.35.

\begin{figure}
\begin{center}
\includegraphics[width=1.0\hsize,angle=0]{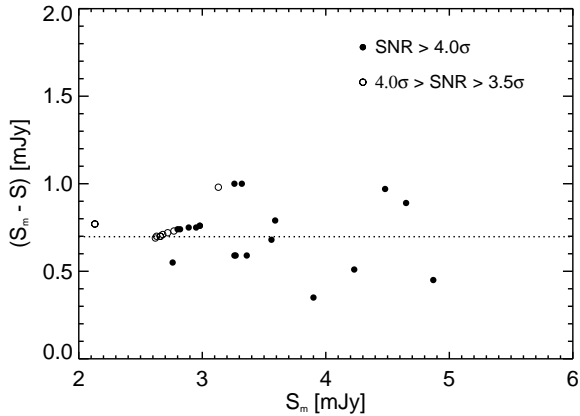}
\caption[]{The difference between the raw and the deboosted
1200-$\mu$m flux densities as a function of the raw flux densities for
sources in the final deboosted catalogue.  On average, the raw flux
densities are reduced by $\sim$0.7\,mJy as a result of the deboosting.}
\label{figure:Sin-Sout}
\end{center}
\end{figure}

Following Coppin et al.\ (2006), only sources with a posterior
probability less than 5 per cent of having a flux density less than or
equal to zero were retained in the final source catalogue. This means
that of the two examples just employed, GN\,1200.3 makes it into the
final catalogue, whereas GN\,1200.35 does not. Adopting this criterion
leads to a total of 17 sources being discarded, all of them originally
detected at below 4\,$\sigma$. The final revised source list is
indicated through column seven of
Table~\ref{table:mambo-source-list}. We adopt the deboosted flux
density as the value at which the $P(S_{\rm p})$ distribution peaks,
with the errors calculated as the 68 per cent confidence bounds on the
non-Gaussian deboosted flux distribution.  On average, the deboosting
reduces the observed fluxes by about 0.7\,mJy, as shown in
Fig.~\ref{figure:Sin-Sout}. In addition, the deboosting results in
larger photometric errors and non-Gaussian probability distributions
(see Fig.~\ref{figure:deboost}), particularly in cases where the
photometric errors are already large.

The final catalogue of deboosted $\ge$3.5-$\sigma$ MAMBO sources
consists of 30 sources.  In \S\ref{subsubsection:spurious-sources} we
found that seven of the 47 $\ge$3.5-$\sigma$ sources recovered prior
to deboosting were spurious. Hence, of the sources in the deboosted
list we expect $\sim$7 $\times$ (30/47), or around four, to be
spurious. This is almost certainly an upper limit since deboosting is
designed to preferentially remove spurious sources.  In
Table~\ref{table:mambo-source-list} we list all of the original
$\ge$3.5-$\sigma$ sources, because many of those which did not pass our
deboosting criteria still have a reasonably high chance of being
genuine and hence may be useful for follow-up studies at other
wavelengths.

%
%
\begin{table*}
\scriptsize
\caption{1200-$\mu$m MAMBO sources extracted from the GOODS-N field at
a significance $\ge$4\,$\sigma$, with a supplementary list of
sources with $3.5\le {\rm S/N}< 4$ (italics). Sources which pass our
deboosting criterion are assigned a deboosted flux density in the 7th
column. These 30 sources make up our robust, deboosted catalogue, of
which we expect no more than 3--4 to be spurious
(\S\ref{subsection:flux-boosting}).  The three last columns list
1.4-GHz flux densities of all robust radio sources within a 6-arcsec
radius of each MAMBO source, their mm-radio positional offset and the
statistical significance of each mm-radio association,
respectively. Associations with $P<0.05$ are considered robust and are
listed in boldface (see also \S~\ref{subsection:mm-radio}).  }
\vspace{0.5cm}
\begin{center}
\begin{tabular}{llccccccccc}
\hline
\hline
Name                 & GN name        &  R.A.\ (J2000)   & Dec.\ (J2000) &   $S_{\rm 1200\mu m} \pm \sigma_{\rm 1200\mu m}$   &   S/N      &   $S_{\rm 1200\mu m}$ (deboosted) & $S_{\rm 1.4GHz}$ & offset  & $P$\\
                   &              &     &    &   /mJy                          &            &      /mJy    & /$\mu$Jy           & /arcsec &\\
\hline
MM\,J123711$+$622211& GN\,1200.1   & 12:37:11.7  & $+$62:22:11  & $9.3\pm 0.9$ & 10.3  & $9.3\pm 0.5$ & $58\pm 16$ & 1.33 & {\bf 0.006}  \\ 
MM\,J123633$+$621407& GN\,1200.2   & 12:36:33.2  & $+$62:14:07  & $4.9\pm 0.7$ & 6.95  & $4.4\pm 0.7$ & $34\pm 4$  & 2.28 & {\bf 0.017}             \\ 
MM\,J123711$+$621328& GN\,1200.3   & 12:37:11.2  & $+$62:13:28  & $3.9\pm 0.6$ & 6.50  & $3.6\pm 0.6$ & $149\pm 14$& 3.16 & {\bf 0.009}\\ 
                    &              &             &              &              &       &              & $62\pm 6$  & 5.92 & {\bf 0.047}\\ 
MM\,J123730$+$621258& GN\,1200.4   & 12:37:30.8  & $+$62:12:59  & $4.2\pm 0.7$ & 6.04  & $3.7\pm 0.7$ & $114\pm 7$ & 0.15 & {\bf 0.00006}\\ 
MM\,J123626$+$620604& GN\,1200.5   & 12:36:26.9  & $+$62:06:04  & $4.7\pm 0.9$ & 5.16  & $3.8\pm 1.0$ & $46\pm 8$  & 3.18 & {\bf 0.025}\\ 
MM\,J123714$+$621827& GN\,1200.6   & 12:37:14.1  & $+$62:18:28  & $3.6\pm 0.7$ & 5.08  & $2.9\pm 0.8$ & $627\pm 8$ & 2.33 & {\bf 0.002}\\ 
MM\,J123607$+$620855& GN\,1200.7   & 12:36:07.8  & $+$62:08:56  & $3.4\pm 0.7$ & 4.80  & $2.8\pm 0.8$ & $\cdots$   & $\cdots$     & $\cdots$          \\ 
MM\,J123702$+$621948& GN\,1200.8   & 12:37:02.9  & $+$62:19:48  & $3.3\pm 0.7$ & 4.67  & $2.7\pm 0.8$ & $\cdots$   & $\cdots$     & $\cdots$           \\ 
MM\,J123644$+$621940& GN\,1200.9   & 12:36:44.6  & $+$62:19:40  & $3.3\pm 0.7$ & 4.65  & $2.7\pm 0.8$ & $\cdots$  & $\cdots$     & $\cdots$\\ 
MM\,J123737$+$621456& GN\,1200.10   & 12:37:37.8  & $+$62:14:56 & $2.8\pm 0.6$ & 4.60  & $2.2\pm 0.7$ & $\cdots$   & $\cdots$     & $\cdots$           \\ 
MM\,J123609$+$620648& GN\,1200.11   & 12:36:09.7  & $+$62:06:48 & $3.6\pm 0.8$ & 4.48  & $2.8\pm 0.9$ & $\cdots$   & $\cdots$     & $\cdots$\\ 
MM\,J123550$+$621041& GN\,1200.12   & 12:35:50.3  & $+$62:10:42 & $4.5\pm 1.0$ & 4.48  & $3.5\pm 1.2$ & $\cdots$   & $\cdots$     &$\cdots$ \\ 
MM\,J123631$+$621715& GN\,1200.13   & 12:36:31.8  & $+$62:17:15 & $3.0\pm 0.7$ & 4.25  & $2.2\pm 0.8$ & $\cdots$   & $\cdots$     & $\cdots$\\ 
MM\,J123635$+$620714& GN\,1200.14   & 12:36:35.6  & $+$62:07:14 & $3.0\pm 0.7$ & 4.25  & $2.2\pm 0.8$ & $\cdots$   & $\cdots$     & $\cdots$\\ 
MM\,J123554$+$621338& GN\,1200.15   & 12:35:54.0  & $+$62:13:38 & $3.0\pm 0.7$ & 4.21  & $2.2\pm 0.9$ & $55\pm 8$  & 5.13 & {\bf 0.043}\\ 
MM\,J123723$+$622049& GN\,1200.16   & 12:37:23.7  & $+$62:20:49 & $3.3\pm 0.8$ & 4.15  & $2.3\pm 1.0$ & $\cdots$   & $\cdots$& $\cdots$\\ 
MM\,J123736$+$621103& GN\,1200.17   & 12:37:36.0  & $+$62:11:04 & $2.9\pm 0.7$ & 4.12  & $2.1\pm 0.9$ & $42\pm 7$  & 5.89 &      0.057\\ 
MM\,J123556$+$621726& GN\,1200.18   & 12:35:56.0  & $+$62:17:26 & $3.3\pm 0.8$ & 4.07  & $2.3\pm 1.0$ & $\cdots$   & $\cdots$& $\cdots$\\ 
MM\,J123645$+$620823& GN\,1200.19   & 12:36:45.1  & $+$62:08:24 & $2.8\pm 0.7$ & 4.02  & $2.1\pm 0.9$ & $\cdots$   & $\cdots$& $\cdots$\\ 
MM\,J123749$+$621558& GN\,1200.20   & 12:37:49.3  & $+$62:15:59 & $2.8\pm 0.7$ & 4.00  & $2.1\pm 0.9$ & $\cdots$   & $\cdots$& $\cdots$\\ 
\hline
\multicolumn{7}{c}{$3.5 \le \sigma < 4$ detections} \\ 
\hline
{\it MM\,J123720$+$621106}& {\it GN\,1200.21}   & {\it 12:37:20.8}  & {\it $+$62:11:06}  & {\it $\rm \it 2.8\pm 0.7$} & {\it 3.95}  & $\rm \it 2.0\pm0.9$ &$\cdots$ & $\cdots$ &$\cdots$\\ 
{\it MM\,J123711$+$620652}& {\it GN\,1200.22}   & {\it 12:37:11.2}  & {\it $+$62:06:52}  & {\it $\rm \it 3.9\pm 1.0$} & {\it 3.92}  & $\cdots$   & $\cdots$& $\cdots$ &$\cdots$\\ 
{\it MM\,J123717$+$620800}& {\it GN\,1200.23}   & {\it 12:37:17.4}  & {\it $+$62:08:00}  & {\it $\rm \it 3.1\pm 0.8$} & {\it 3.91}  & $\rm \it 2.2\pm1.1$ & $97\pm 19$ & 4.24 & {\bf 0.021}\\ 
{\it MM\,J123755$+$621057}& {\it GN\,1200.24}   & {\it 12:37:55.6}  & {\it $+$62:10:58}  & {\it $\rm \it 3.5\pm 0.9$} & {\it 3.91}  & $\cdots$  & $\cdots$& $\cdots$ &$\cdots$\\ 
{\it MM\,J123650$+$621217}& {\it GN\,1200.25}   & {\it 12:36:50.9}  & {\it $+$62:12:17}  & {\it $\rm \it 2.7\pm 0.7$} & {\it 3.88}  & $\rm \it 2.0\pm0.9$ & $\cdots$ & $\cdots$ & $\cdots$\\ 
{\it MM\,J123643$+$622350}& {\it GN\,1200.26}   & {\it 12:36:43.9}  & {\it $+$62:23:51}  & {\it $\rm \it 4.7\pm 1.2$} & {\it 3.87}  & $\cdots$  & $\cdots$& $\cdots$ &$\cdots$\\ 
{\it MM\,J123607$+$621137}& {\it GN\,1200.27}   & {\it 12:36:07.2}  & {\it $+$62:11:37}  & {\it $\rm \it 2.7\pm 0.7$} & {\it 3.82}  & $\rm \it 2.0\pm1.0$ & $\cdots$ &$\cdots$ &$\cdots$\\ 
{\it MM\,J123605$+$621650}& {\it GN\,1200.28}   & {\it 12:36:05.8}  & {\it $+$62:16:50}  & {\it $\rm \it 2.7\pm 0.7$} & {\it 3.80}  & $\rm \it 2.0\pm1.0$ & $\cdots$ &$\cdots$ &$\cdots$\\ 
{\it MM\,J123711$+$621218}& {\it GN\,1200.29}   & {\it 12:37:11.8}  & {\it $+$62:12:18}  & {\it $\rm \it 2.7\pm 0.7$} & {\it 3.80}  & $\rm \it 2.0\pm1.0$ & $\cdots$ &$\cdots$ &$\cdots$\\ 
{\it MM\,J123638$+$622049}& {\it GN\,1200.30}   & {\it 12:36:38.8}  & {\it $+$62:20:50}  & {\it $\rm \it 3.0\pm 0.8$} & {\it 3.78}  & $\cdots$  & $\cdots$& $\cdots$ &$\cdots$\\ 
{\it MM\,J123803$+$621446}& {\it GN\,1200.31}   & {\it 12:38:03.4}  & {\it $+$62:14:46}  & {\it $\rm \it 3.0\pm 0.8$} & {\it 3.77}  & $\cdots$  & $\cdots$& $\cdots$ &$\cdots$\\ 
{\it MM\,J123711$+$621924}& {\it GN\,1200.32}   & {\it 12:37:11.8}  & {\it $+$62:19:24}  & {\it $\rm \it 2.6\pm 0.7$} & {\it 3.75}  & $\rm \it 1.9\pm1.0$ & $\cdots$ &$\cdots$ &$\cdots$\\ 
{\it MM\,J123641$+$621129}& {\it GN\,1200.33}   & {\it 12:36:41.9}  & {\it $+$62:11:30}  & {\it $\rm \it 2.6\pm 0.7$} & {\it 3.74}  & $\rm \it 1.9\pm1.0$ & $\cdots$ &$\cdots$ &$\cdots$\\ 
{\it MM\,J123752$+$621428}& {\it GN\,1200.34}   & {\it 12:37:52.0}  & {\it $+$62:14:28}  & {\it $\rm \it 2.6\pm 0.7$} & {\it 3.70}  & $\cdots$  & $\cdots$& $\cdots$ &$\cdots$\\ 
{\it MM\,J123541$+$621150}& {\it GN\,1200.35}   & {\it 12:35:41.7}  & {\it $+$62:11:50}  & {\it $\rm \it 4.1\pm 1.1$} & {\it 3.69}  & $\cdots$  &$\cdots$ & $\cdots$ &$\cdots$\\ 
{\it MM\,J123802$+$621940}& {\it GN\,1200.36}   & {\it 12:38:02.8}  & {\it $+$62:19:40}  & {\it $\rm \it 4.4\pm 1.2$} & {\it 3.68}  & $\cdots$  &$\cdots$ & $\cdots$ &$\cdots$\\ 
{\it MM\,J123656$+$621647}& {\it GN\,1200.37}   & {\it 12:36:56.8}  & {\it $+$62:16:47}  & {\it $\rm \it 2.6\pm 0.7$} & {\it 3.65}  & $\cdots$  &$\cdots$ & $\cdots$ &$\cdots$\\ 
{\it MM\,J123540$+$621246}& {\it GN\,1200.38}   & {\it 12:35:40.3}  & {\it $+$62:12:47}  & {\it $\rm \it 3.7\pm 1.0$} & {\it 3.65}  & $\cdots$  &$\cdots$ & $\cdots$ &$\cdots$\\ 
{\it MM\,J123714$+$621842}& {\it GN\,1200.39}   & {\it 12:37:14.1}  & {\it $+$62:18:43}  & {\it $\rm \it 2.6\pm 0.7$} & {\it 3.64}  & $\cdots$  &$\cdots$ & $\cdots$ &$\cdots$\\ 
{\it MM\,J123745$+$621129}& {\it GN\,1200.40}   & {\it 12:37:45.4}  & {\it $+$62:11:30}  & {\it $\rm \it 2.5\pm 0.7$} & {\it 3.61}  & $\cdots$  &$\cdots$ & $\cdots$ &$\cdots$\\ 
{\it MM\,J123621$+$621705}& {\it GN\,1200.41}   & {\it 12:36:21.5}  & {\it $+$62:17:06}  & {\it $\rm \it 2.5\pm 0.7$} & {\it 3.55}  & $\cdots$  & $138\pm 7$ &  2.86 & {\bf 0.008}\\ 
                          &                     &                   &                    &                            &             &           & $50\pm 7$ & 5.24 & {\bf 0.046}\\ 
{\it MM\,J123700$+$621453}& {\it GN\,1200.42}   & {\it 12:37:00.6}  & {\it $+$62:14:54}  & {\it $\rm \it 2.1\pm 0.6$} & {\it 3.55}  & $\rm \it 1.4\pm0.8$ & $53\pm 13$ & 5.84 & 0.051\\ 
{\it MM\,J123712$+$621258}& {\it GN\,1200.43}   & {\it 12:37:12.7}  & {\it $+$62:12:59}  & {\it $\rm \it 2.1\pm 0.6$} & {\it 3.55}  & $\rm \it 1.4\pm0.8$ &$\cdots$ & $\cdot$ &$\cdots$\\ 
{\it MM\,J123659$+$621619}& {\it GN\,1200.44}   & {\it 12:36:59.2}  & {\it $+$62:16:19}  & {\it $\rm \it 2.5\pm 0.7$} & {\it 3.54}  & $\cdots$  &$\cdots$ & $\cdots$ &$\cdots$\\ 
{\it MM\,J123653$+$622401}& {\it GN\,1200.45}   & {\it 12:36:53.4}  & {\it $+$62:24:01}  & {\it $\rm \it 4.2\pm 1.2$} & {\it 3.53}  & $\cdots$  &$\cdots$ & $\cdots$ &$\cdots$\\ 
{\it MM\,J123655$+$621408}& {\it GN\,1200.46}   & {\it 12:36:55.5}  & {\it $+$62:14:09}  & {\it $\rm \it 2.1\pm 0.6$} & {\it 3.51}  & $\cdots$  &$\cdots$ & $\cdots$ &$\cdots$\\ 
{\it MM\,J123557$+$621250}& {\it GN\,1200.47}   & {\it 12:35:57.6}  & {\it $+$62:12:50}  & {\it $\rm \it 2.5\pm 0.7$} & {\it 3.51}  & $\cdots$  &$\cdots$ & $\cdots$ &$\cdots$\\ 
\hline
\label{table:mambo-source-list}
\end{tabular}
\end{center}
\end{table*}

\section{Comparison with the SCUBA GOODS-N survey}
\label{section:mambo-scuba-comparison}

The SCUBA 850-$\mu$m survey of the Hubble Deep Field North (HDF-N),
which lies at the centre of the GOODS-N field, was one of the first
extragalactic fields to be observed at submm wavelengths (Hughes et
al.\ 1998; Serjeant et al.\ 2003) and to this day remains one of the
deepest blank-field submm maps.  A much larger, shallower 850-$\mu$m
image, covering most of the GOODS-N field, was presented by Borys et
al.\ (2002) and this has been supplemented by several SCUBA jiggle
maps (Holland et al.\ 1999) of other parts of GOODS-N (Borys et al.\
2003, 2004; Wang et al.\ 2004; Pope et al.\ 2005; Pope et al.\
2006). These data were combined into the so-called `super-map' of
GOODS-N by Borys et al.\ (2003), then updated in a series of
subsequent papers. The resulting map combined the available data
optimally, with the purpose of extracting point sources. The final
version of the 850-$\mu$m source catalogue was presented by Pope et
al.\ (2006), with three new sources (GN\,850.39, GN\,850.40 and
GN\,850.41) presented by Wall, Pope \& Scott (2008).

The AzTEC camera (Wilson et al.\ 2008) was used to survey the GOODS-N region at 1100-$\mu$m
(Perera et al.\ 2008). A comparison of the MAMBO and AzTEC millimetre maps will
be explored in a future paper (Pope et al.\ in preparation).

In this section we will compare the 850-$\mu$m super-map with our new
1200-$\mu$m image, with two goals: first, to extract a robust sample
of SMGs detected jointly at 850 and 1200\,$\mu$m; second, to compare
the degree of overlap between the populations selected at each
wavelength.

The optimal method for extracting sources {\it of arbitrary spectral
shape} is to extract sources from the combined S/N map, thereby using
all the information from the 1200- and 850-$\mu$m data. However, some
care must be exercised, for at least two reasons.  First, one is
digging into the confusion noise in both images and hence
non-Gaussianity or systematic artefacts in the maps may start to have
effects that were not previously evident. Second, if one had strong
prior information about the expected $\fdr$ ratio for the emitters
then one would adopt a different weighting scheme.

%
%

The source-extraction method applied to the two maps individually (see
Borys et al.\ 2003) is optimal, with the caveats given in
\S~\ref{subsection:source-extraction} for finding point sources, under
the assumption that the noise distributions in both maps are Gaussian
and uncorrelated. The SCUBA super-map can be thought of as the answer
to the question: `what is the best estimate of the flux density of a
point source centred on each pixel?'. The super-map combines data
taken in different observing modes, with different chop-throws, etc.,
and also takes into account the beam convolution in the process of
fitting to the PSF. It follows that an efficient way of extracting
point sources, using the information in both maps, is simply to treat
the MAMBO image as additional data. These can then be combined with the
SCUBA data to provide an overall S/N `map', which gives the combined
S/N level for a point source centred on each pixel. The relevant
expression for each pixel is:
\begin{equation}
S = \frac{S_{\rm M}/\sigma^2_{\rm M} +
S_{\rm S}/\sigma^2_{\rm S}}{1/\sigma^2_{\rm M}
+ 1/\sigma^2_{\rm S}}, 
\label{equation:combo-eq-1}
\end{equation}
where $\sigma_{\rm M}$ and $\sigma_{\rm S}$ are the MAMBO and SCUBA
noise maps, respectively. The corresponding noise map is
\begin{equation}
\sigma = \sqrt{\sigma^2_{\rm M} + \sigma^2_{\rm S}}. 
\label{equation:combo-eq-2}
\end{equation}
For this method to work straightforwardly, the two maps need to be
astrometrically aligned and to have the same pixel scale and
resolution. To this end, both maps were binned onto a 1-arcsec$^2$
pixel grid and we checked that no shift between the grids was required
in order to maximise the correlation, then the MAMBO image was
smoothed to the SCUBA resolution. We stress that this method only
improves on the robustness of our detections, it does not provide
estimates of the 850- and 1200-$\mu$m flux densities, which have to be
measured from the separate maps.

\begin{figure*}
\begin{center}
\includegraphics[width=0.63\hsize,angle=-90]{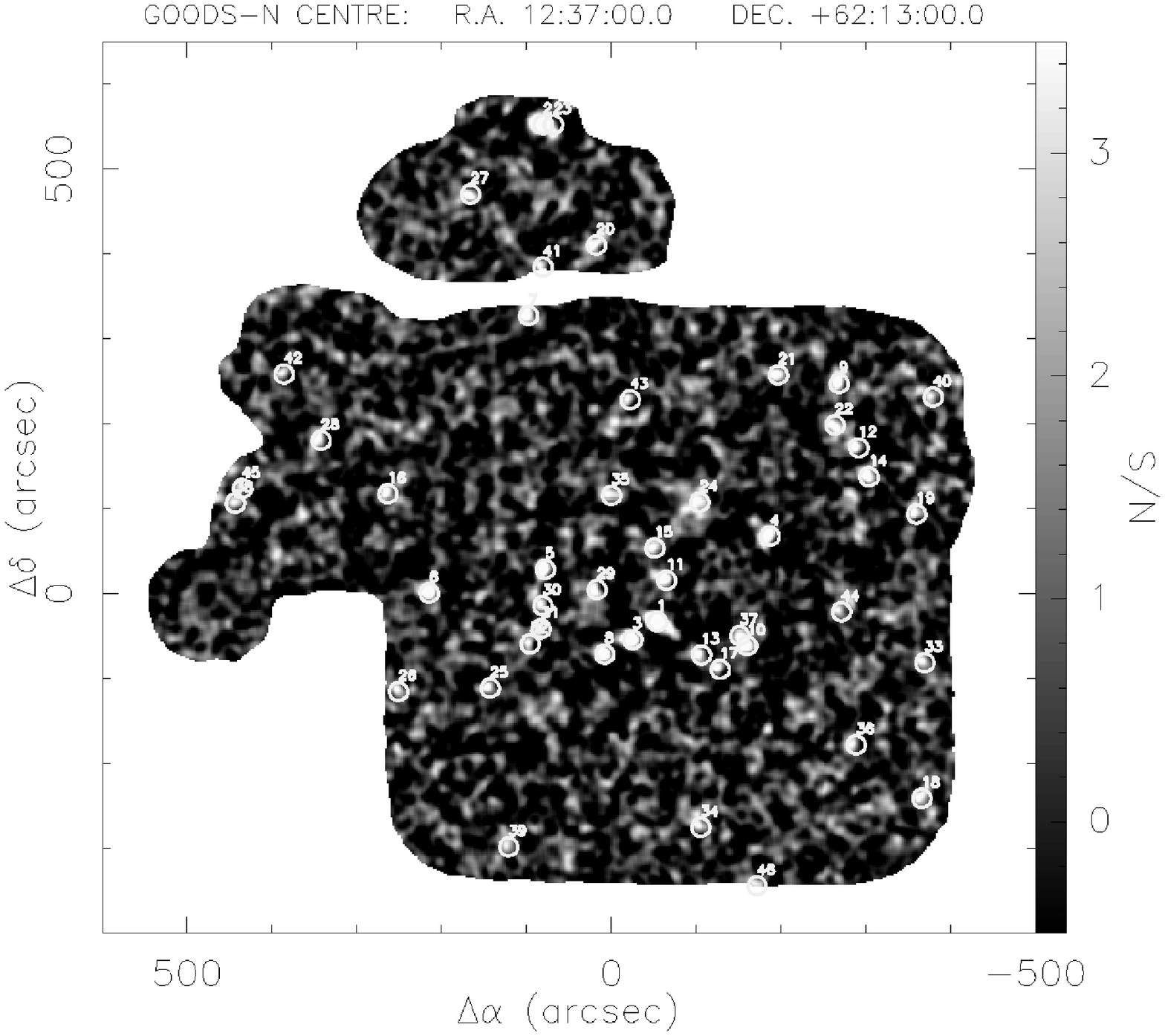}
\includegraphics[width=0.63\hsize,angle=-90]{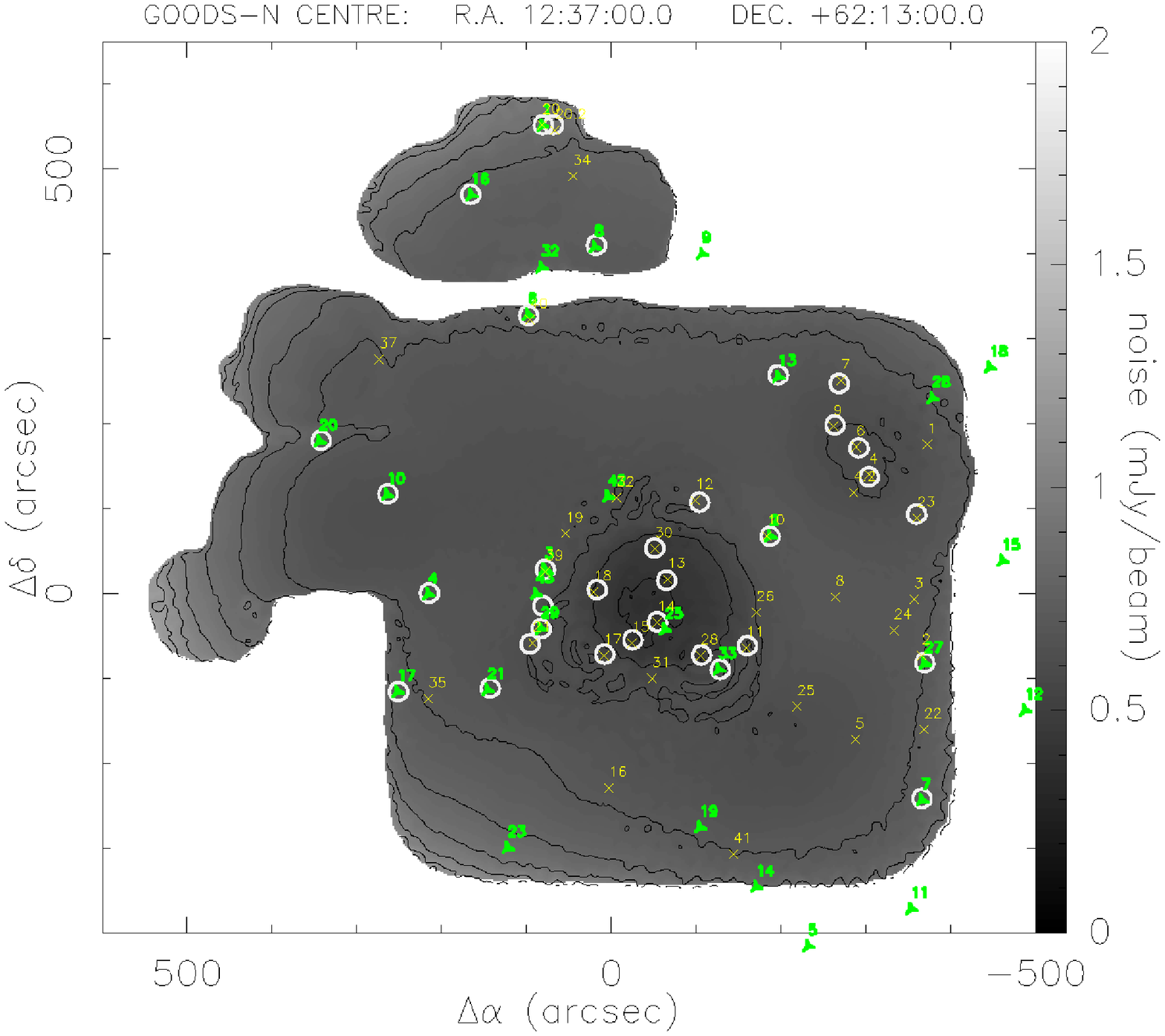}
\caption[]{S/N image (top) and r.m.s.\ noise map (bottom) resulting
from combining the 850-$\mu$m SCUBA map (Borys et al.\ 2003; Pope et
al.\ 2005) and 1200-$\mu$m MAMBO map (this work) according to
eqs~\ref{equation:combo-eq-1} and \ref{equation:combo-eq-2}. White
circles indicate sources with $\ge$4-$\sigma$ significance from the
combined image. For the sake of clarity, these sources have not been
numbered in the bottom panel. Green triangles indicate the deboosted
$\ge$3.5-$\sigma$ MAMBO sources in
Table~\ref{table:mambo-source-list}; yellow crosses are the deboosted
$\ge$3.5-$\sigma$ SCUBA sources, as given by Pope et al.\ (2006), with
the addition of GN\,850.39, .40 and .41 from Wall, Pope \& Scott
(2008).}
\label{figure:mambo-scuba-maps}
\end{center}
\end{figure*}

Fig.~\ref{figure:mambo-scuba-maps} shows the combined 850- and
1200-$\mu$m S/N and r.m.s.\ noise maps obtained in this way, and
Table~\ref{table:mambo-scuba-source-list} lists the sources extracted
from the combined map.

A total of 13 (33) sources are detected above 5\,$\sigma$
(4\,$\sigma$) in the combined map. In the bottom panel of
Fig.~\ref{figure:mambo-scuba-maps}, and in
Table~\ref{table:mambo-scuba-source-list}, we compare our dual-map
source list with the SCUBA-only and MAMBO-only source lists.  Any
source from the combined map which has a $\ge$3.5-$\sigma$ deboosted
MAMBO (SCUBA) source within 10\,arcsec ($\sqrt{8^2 + 6^2}$\,arcsec,
where 8 and 6\,arcsec are the 95-per-cent confidence limit for SCUBA
and MAMBO sources (Ivison et al.\ 2005; Pope et al.\ 2006) is deemed
to have been {\em associated} at 1200\,$\mu$m (850\,$\mu$m).

How many sources have robust detections at both 850 and 1200\,$\mu$m?
Only six sources from the combined list, four of which are detected at
$\ge$5\,$\sigma$ in the combined map, and one at 4\,$\sigma$. The sixth source, GN\,1200.42
(GN\,850.32), is detected at $\sim$3.8\,$\sigma$. In cases with no
detection at 850 or 1200\,$\mu$m, we have determined 3-$\sigma$ upper
flux density limits (Table~\ref{table:mambo-scuba-source-list}).
Finally, we note that one source detected at $\ge$4\,$\sigma$ in the
combined map, MM\,J123711$+$621245, was not previously known.

%
%

How big a fraction of the MAMBO and SCUBA sources are recovered in the
combined map?  All five (14) of the $\ge$5-$\sigma$ ($\ge$4-$\sigma$)
MAMBO sources which lie within the SCUBA region are recovered in the
combined map.  Conversely, all of the $\ge$5-$\sigma$ SCUBA sources
are detected in the combined map, while for the $\ge$4-$\sigma$ SCUBA
sources the corresponding fraction is 18/24 (75 per cent).

If we instead use Table~\ref{table:mambo-scuba-source-list} to answer
the question `how big a fraction of the MAMBO sources have robust
associations in the SCUBA map?', we find that apart from the four
$\ge$5-$\sigma$ MAMBO sources with SCUBA counterparts (a 100-per-cent
recovery rate since the fifth $\ge$5-$\sigma$ MAMBO source,
GN\,1200.5, lies outside the SCUBA region) none of the other
$\ge$4-$\sigma$ MAMBO sources have been recovered at 850\,$\mu$m.
This is clearly at odds with the high fraction of associations found
for $\ge$4-$\sigma$ MAMBO sources by G04 in ELAIS\,N2 and Lockman Hole
($\sim$30--50 per cent) and the similarly high rate of associations
for bright 1.1-mm BOLOCAM sources in the Lockman Hole ($\sim$75 per
cent -- Laurent et al.\ 2006).  While positional uncertainties may go
some way towards explaining the low fraction of associations in
GOODS-N, the dominant factor is probably the mis-match in sensitivity
of the MAMBO and SCUBA surveys in this region.

Finally, we note that GN\,1200.4 -- one of the brightest ($S_{\rm
1200\mu m} = 4.2\pm 0.7$\,mJy) and most significant ($\ge$5\,$\sigma$)
sources in our sample -- is not detected at 850\,$\mu$m and is thus a
submm drop-out (SDO) candidate, although it does lie in a relatively noisy part of the
SCUBA map.

%
%
\begin{table*}
\scriptsize
\caption{Sources extracted at $\ge$4\,$\sigma$ from the combined 1200-
and 850-$\mu$m image (eqs~\ref{equation:combo-eq-1} and
\ref{equation:combo-eq-2}), with a supplementary list of sources
extracted at $3.5\le {\rm S/N} < 4$ (italics).  The indices in
parentheses refer to the numbering of the MAMBO catalogue in Table 1,
except when there is no MAMBO association in which case we have
started the index at 100. The same numbering is used in
Fig.~\ref{figure:mambo-scuba-maps}. The MAMBO and SCUBA associations
in the second column from the left are those given in
Table~\ref{table:mambo-source-list} and in Pope et al.\ (2005),
respectively. Deboosted 1200- and 850-$\mu$m flux densities are given
in parentheses in columns 6 and 7. In cases where no robust
associations have been identified, we list the 3-$\sigma$ upper flux
density limit. Columns 8, 9 and 10 list $S_{\rm 1.4GHz}$ for all
robust radio identifications within 8\,arcsec of each MAMBO/SCUBA
association in the combined map, the radial offset and the statistical
significance of each radio identification,
respectively. Identifications with $P<0.05$ are considered robust and
are given in boldface.}
\vspace{0.5cm}
\begin{center}
\begin{tabular}{rlcccccccc}
\hline
\hline
\multicolumn{1}{l}{Name}    &  GN name                &R.A.\ (J2000)   & Dec.\ (J2000)  & S/N &  $S_{\rm 1200\mu m} \pm \sigma_{\rm 1200\mu m}$   &   $S_{\rm 850\mu m}\pm \sigma_{\rm 850\mu m}$     &  $S_{\rm 1.4GHz}$  & offset & $P$  \\
                            &                          &   &      &           &  /mJy                  &   /mJy                                                       &   /$\mu$Jy          &      \\
\hline
(100) MM\,J123652$+$621226  &  GN\,850.14              & 12:36:52.2  & $+$62:12:26  & 14.7      &  $\le 2.1$            &   $5.9\pm 0.3$ ($5.9$)   &  $< 16$           & $\cdots$  & $\cdots$  \\
  (1) MM\,J123711$+$622212  &  GN\,1200.1/GN\,850.20   & 12:37:11.5  & $+$62:22:12  & 13.5      &  $9.3\pm 0.9$ ($9.3$) &   $20.3\pm 2.1$ ($20.3$) &  $58\pm 16$       & 2.63      & {\bf 0.016 }\\
(101) MM\,J123656$+$621205  &  GN\,850.15              & 12:36:56.4  & $+$62:12:05  & 8.75      &  $\le 2.0$            &   $3.7\pm 0.4$  ($3.7$)  &  $42\pm 6$        & 3.02      & {\bf 0.024}      \\
  (2) MM\,J123633$+$621407  &  GN\,1200.2/GN\,850.10   & 12:36:33.2  & $+$62:14:07  & 7.85      &  $4.9\pm 0.7$ ($4.4$) &   $11.3\pm 1.6$ ($11.3$) &  $34\pm 4$        & 2.28      & {\bf 0.017} \\
  (3) MM\,J123711$+$621328  &  GN\,1200.3/GN\,850.39   & 12:37:11.1  & $+$62:13:28  & 7.16      &  $3.9\pm 0.6$ ($3.6$) &   $7.4\pm 2.0$ ($5.2$)   &  $149\pm 14$      & 3.42      & {\bf 0.010}      \\
                            &                          &             &              &           &                       &                          &  $62\pm 6$        & 6.57      &      0.053      \\
  (4) MM\,J123730$+$621300  &  GN\,1200.4              & 12:37:30.7  & $+$62:13:00  & 6.14      &  $4.2\pm 0.7$ ($3.7$) &   $\le 15.6$             &  $114\pm 7$       & 1.23      & {\bf 0.002}  \\
  (6) MM\,J123714$+$621827  &  GN\,1200.6/GN\,850.40   & 12:37:14.0  & $+$62:18:27  & 6.00      &  $3.6\pm 0.7$ ($2.9$) &   $13.1\pm 2.7$ ($10.7$) &  $627\pm 7$       & 1.16      & {\bf 0.0005}        \\
(102) MM\,J123701$+$621148  &  GN\,850.17              & 12:37:01.1  & $+$62:11:48  & 6.00      &  $\le 2.0$            &   $3.9\pm 0.7$ ($3.9$)   &  $102\pm 12$      & 3.31      & {\bf 0.014}\\
(103) MM\,J123621$+$621707  &  GN\,850.7               & 12:36:21.5  & $+$62:17:07  & 5.83      &  $\le 2.0$            &   $8.9\pm 1.5$ ($8.9$)   &  $138\pm 7$       & 2.09      & {\bf 0.005}\\
                            &                          &             &              &           &                       &                          &                   & 4.59      & {\bf 0.039}\\
(104) MM\,J123637$+$621158  &  GN\,850.11              & 12:36:37.1  & $+$62:11:58  & 5.83      &  $\le 2.0$            &   $7.0\pm 0.9$ ($7.0$)   &  $26\pm 5$        & 3.19      & {\bf 0.028}\\
(105) MM\,J123650$+$621315  &  GN\,850.13              & 12:36:50.7  & $+$62:13:15  & 5.25      &  $\le 2.1$            &   $1.9\pm 0.4$ ($1.9$)   &  $47\pm 6$        & 7.27      &      0.068\\
(106) MM\,J123618$+$621551  &  GN\,850.6               & 12:36:18.2  & $+$62:15:51  & 5.00      &  $\le 2.0$            &   $7.5\pm 0.9$ ($7.5$)   &  $170\pm 7$       & 1.08      & {\bf 0.001}\\
(107) MM\,J123644$+$621147  &  GN\,850.28              & 12:36:44.8  & $+$62:11:47  & 5.00      &  $\le 2.0$            &   $1.7\pm 0.4$ ($1.7$)   &  $90\pm 24$       & 5.38      & {\bf 0.031}      \\
(108) MM\,J123616$+$621517  &  GN\,850.4               & 12:36:16.5  & $+$62:15:17  & 4.83      &  $\le 2.0$            &   $5.1\pm 1.0$ ($4.9$)   &  $58\pm 7$        & 4.13      & {\bf 0.031}\\
(109) MM\,J123652$+$621353  &  GN\,850.30              & 12:36:52.7  & $+$62:13:53  & 4.75      &  $\le 2.0$            &   $1.8\pm 0.5$ ($1.8$)   &  $20\pm 5$        & 1.82      & {\bf 0.011}\\
 (10) MM\,J123737$+$621457  &  GN\,1200.10             & 12:37:37.7  & $+$62:14:57  & 4.66      &  $2.8\pm 0.6$ ($2.2$) &   $\le 6.0$              &  $\cdots$         & $\cdots$  & $\cdots$\\
 (33) MM\,J123641$+$621130  &  GN\,1200.33             & 12:36:41.6  & $+$62:11:30  & 4.60      &  $2.6\pm 0.7$ ($1.9$) &   $\le 3.0$              &  $\cdots$         & $\cdots$  & $\cdots$\\
  (7) MM\,J123607$+$620858  &  GN\,1200.7              & 12:36:07.8  & $+$62:08:58  & 4.57      &  $3.4\pm 0.7$ ($2.8$) &   $\le 19.4$             &  $\cdots$         & $\cdots$  & $\cdots$\\
(110) MM\,J123608$+$621433  &  GN\,850.23              & 12:36:08.5  & $+$62:14:33  & 4.50      &  $\le 2.0$            &   $7.0\pm 1.9$ ($4.9$)   &  $48\pm 12$       & 2.45      & {\bf 0.017} \\
  (8) MM\,J123702$+$621950  &  GN\,1200.8              & 12:37:02.6  & $+$62:19:50  & 4.42      &  $3.3\pm 0.7$ ($2.7$) &   $\le 10.7$             &  $\cdots$         & $\cdots$  & $\cdots$\\
 (13) MM\,J123631$+$621717  &  GN\,1200.13             & 12:36:31.8  & $+$62:17:17  & 4.42      &  $3.0\pm 0.7$ ($2.2$) &   $\le 15.5$             &  $\cdots$         & $\cdots$  & $\cdots$\\
(111) MM\,J123622$+$621618  &  GN\,850.9               & 12:36:22.2  & $+$62:16:18  & 4.33      &  $\le 2.0$            &   $8.9\pm 1.0$ ($8.9$)   &  $< 19.5$         & $\cdots$  & $\cdots$\\
(112) MM\,J123709$+$622212  &  GN\,850.20.2            & 12:37:09.9  & $+$62:22:12  & 4.22      &  $\le 2.7$            &   $11.7\pm 2.2$ ($9.9$)  &  $160\pm 9$       & 6,45      & {\bf 0.024}\\
(113) MM\,J123645$+$621447  &  GN\,850.12              & 12:36:45.1  & $+$62:14:47  & 4.16      &  $\le 2.0$            &   $8.6\pm 1.4$ ($8.6$)   &  $103\pm 6$       & 6.87      & {\bf 0.038}\\
 (21) MM\,J123720$+$621108  &  GN\,1200.21             & 12:37:20.4  & $+$62:11:08  & 4.14      &  $2.8\pm 0.7$ ($2.0$) &   $\le 10.0$             &  $\cdots$         & $\cdots$  & $\cdots$\\
 (17) MM\,J123735$+$621104  &  GN\,1200.17             & 12:37:35.8  & $+$62:11:04  & 4.14      &  $2.9\pm 0.7$ ($2.1$) &   $\le 3.0$             &  $42\pm 7$        & 7.0       &      0.061\\
 (16) MM\,J123723$+$622050  &  GN\,1200.16             & 12:37:23.8  & $+$62:20:50  & 4.00      &  $3.3\pm 0.8$ ($2.3$) &   $\le 4.7$              &  $\cdots$         & $\cdots$  & $\cdots$\\
 (20) MM\,J123749$+$621560  &  GN\,1200.20             & 12:37:49.0  & $+$62:16:00  & 4.00      &  $2.8\pm 0.7$ ($2.1$) &   $\le 23.9$             &  $\cdots$         & $\cdots$  & $\cdots$\\
(114) MM\,J123702$+$621304  &  GN\,850.18              & 12:37:02.4  & $+$62:13:04  & 4.00      &  $\le 2.0$            &   $3.2\pm 0.6$ ($3.2$)   &  $21\pm 6$        & 1.95      & {\bf 0.013}\\
(115) MM\,J123711$+$621245  &  $\cdots$                & 12:37:11.5  & $+$62:12:45  & 4.00      &  $\le 2.0$            &   $\le 4.8$              &  $\cdots$         & $\cdots$  & $\cdots$\\
 (29) MM\,J123711$+$621218  &  GN\,1200.29             & 12:37:11.8  & $+$62:12:18  & 4.00      &  $2.7\pm 0.6$ ($2.0$) &   $\le 4.0$              &  $\cdots$         & $\cdots$  & $\cdots$\\
(116) MM\,J123713$+$621200  &  GN\,850.21              & 12:37:13.7  & $+$62:12:00  & 4.00      &  $\le 2.0$            &   $5.7\pm 1.2$ ($4.9$)   &  $34\pm 6$        & 4.12      & {\bf 0.040}\\
 (27) MM\,J123607$+$621138  &  GN\,1200.27/GN\,850.2   & 12:36:07.2  & $+$62:11:38  & 4.00      &  $2.7\pm 0.7$ ($2.0$) &   $16.2\pm4.1$ ($12.1$)  &  $\cdots$         & $\cdots$  & $\cdots$\\
\hline
\multicolumn{7}{c}{{\it $\rm \it 3.5 \le \sigma < 4$ detections}} \\ 
\hline
{\it ~~(19) MM\,J123645$+$620824} & {\it GN\,1200.19}            & {\it 12:36:44.9} & {\it $+$62:08:24} & {\it 3.85} &  {\it $\rm \it 2.8\pm 0.7$ ($2.1$)} & $\rm \it \le 7.3$            & $\cdots$       & $\cdots$    & $\cdots$    \\
{\it ~~(42) MM\,J123700$+$621455} & {\it GN\,1200.42/GN\,850.32} & {\it 12:37:00.0} & {\it $+$62:14:55} & {\it 3.83} &  {\it $\rm \it 2.1\pm 0.6$ ($1.4$)} & $\rm \it 5.3\pm 1.4$ ($3.8$) & $53\pm 13$     & 5.04        & {\bf 0.043} \\
{\it (117) MM\,J123618$+$621001}  & {\it GN\,850.5}              & {\it 12:36:18.8} & {\it $+$62:10:01} & {\it 3.83} &  {\it $\rm \it \le 1.9$}            & $\rm \it 6.7\pm 1.6$ ($5.2$) & $50\pm 15$     & 4.14        & {\bf 0.034} \\
{\it (118) MM\,J123638$+$621210}  & {\it $\cdots$}               & {\it 12:36:38.4} & {\it $+$62:12:10} & {\it 3.80} &  {\it $\rm \it \le 2.0$}            & $\rm \it \le 5.9$            & $\cdots$       & $\cdots$    & $\cdots$    \\
{\it (119) MM\,J123803$+$621446}  & {\it $\cdots$}               & {\it 12:38:03.5} & {\it $+$62:14:46} & {\it 3.75} &  {\it $\rm \it \le 2.1$}            & $\rm \it \le 2.5$            & $\cdots$       & $\cdots$    & $\cdots$    \\
{\it ~~(23) MM\,J123717$+$620801} & {\it GN\,1200.23}            & {\it 12:37:17.3} & {\it $+$62:08:01} & {\it 3.75} &  {\it $\rm \it 3.1\pm 0.8$ ($2.2$)} & $\rm \it \le 3.0$            & $97\pm 19$     & 3.38        & {\bf 0.015} \\
{\it ~~(28) MM\,J123605$+$621651} & {\it GN\,1200.28}            & {\it 12:36:05.6} & {\it $+$62:16:51} & {\it 3.71} &  {\it $\rm \it 2.7\pm 0.7$ ($2.0$)} & $\rm \it \le 16.9$           & $\cdots$       & $\cdots$    & $\cdots$    \\
{\it ~~(32) MM\,J123711$+$621924} & {\it GN\,1200.32}            & {\it 12:37:11.5} & {\it $+$62:19:24} & {\it 3.62} &  {\it $\rm \it 2.6\pm 0.7$ ($1.9$)} & $\rm \it \le 10.9$           & $\cdots$       & $\cdots$    & $\cdots$    \\
{\it (120) MM\,J123755$+$621718}  & {\it $\cdots$}               & {\it 12:37:55.3} & {\it $+$62:17:18} & {\it 3.62} &  {\it $\rm \it \le 2.1$}            & $\rm \it \le 46.4$           & $\cdots$       & $\cdots$    & $\cdots$    \\
{\it (121) MM\,J123656$+$621648}  & {\it $\cdots$}               & {\it 12:36:56.8} & {\it $+$62:16:48} & {\it 3.57} &  {\it $\rm \it \le 2.6$}            & $\rm \it \le 4.6$            & $\cdots$       & $\cdots$    & $\cdots$    \\
{\it (122) MM\,J123621$+$621238}  & {\it $\cdots$}               & {\it 12:36:21.2} & {\it $+$62:12:38} & {\it 3.57} &  {\it $\rm \it \le 2.0$}            & $\rm \it \le 14.5$           & $\cdots$       & $\cdots$    & $\cdots$    \\
{\it (123) MM\,J123802$+$621503}  & {\it $\cdots$}               & {\it 12:38:02.4} & {\it $+$62:15:03} & {\it 3.50} &  {\it $\rm \it \le 2.0$}            & $\rm \it \le 15.1$           & $\cdots$       & $\cdots$    & $\cdots$    \\
{\it ~~(14) MM\,J123635$+$620715} & {\it GN\,1200.14}            & {\it 12:36:35.5} & {\it $+$62:07:15} & {\it 3.50} &  {\it $\rm \it 3.0\pm 0.7$ ($2.2$)} & $\rm \it \le 19.5$           & $42\pm 13$     & 7.60        &      0.073  \\
\hline
\label{table:mambo-scuba-source-list}
\end{tabular}
\end{center}
\end{table*}

\section{The 850-/1200-micron flux density ratio}
\label{section:scuba-mambo-flux-ratio}

In this section we shall examine the 850-$\mu$m to 1200-$\mu$m flux density
($\fdr$) distribution of the (sub)mm sources in GOODS-N.

\subsection{Sources detected by both SCUBA and MAMBO}

\begin{figure}
\begin{center}
\includegraphics[width=1.0\hsize,angle=0]{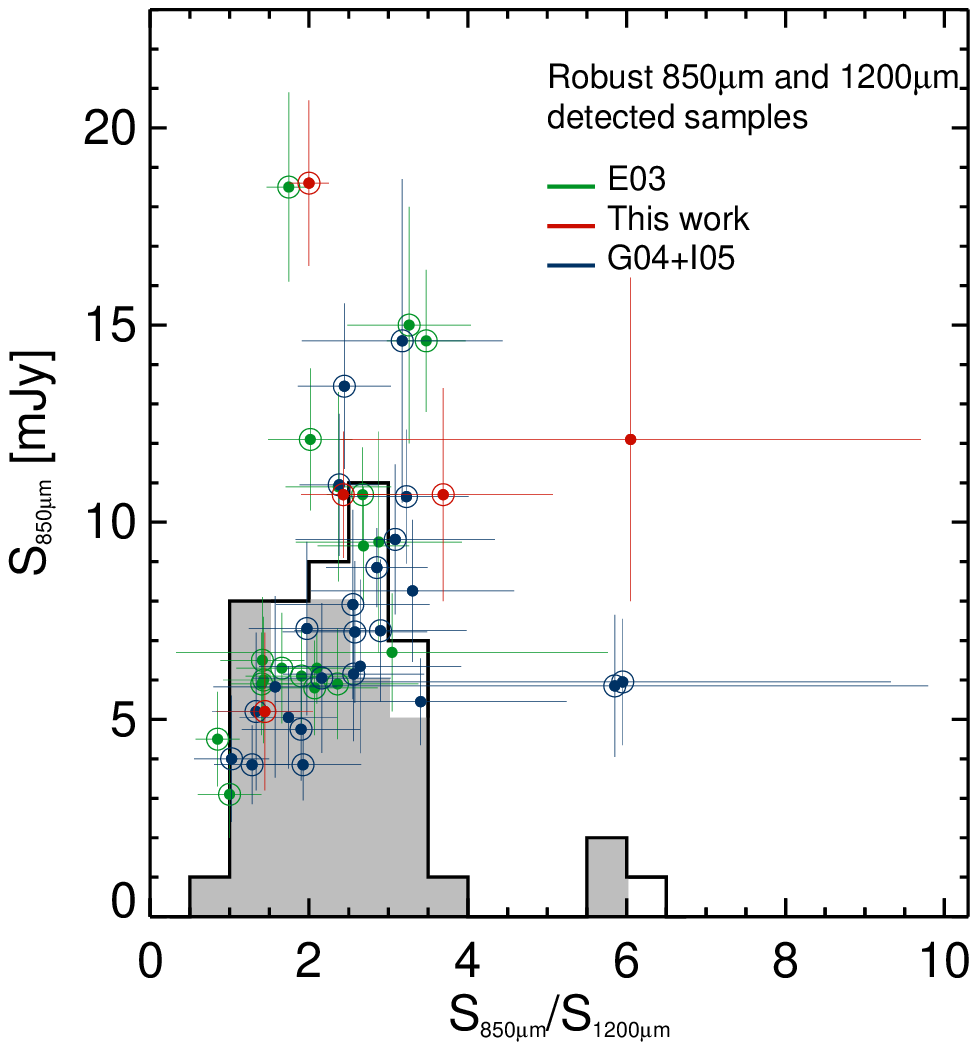}
\caption[]{$S_{\rm 850\mu m}$ versus the $\fdr$ ratio, with corresponding
uncertainties, for the deboosted sources identified in both the SCUBA
{\em and} MAMBO source catalogues of the GOODS-N field (red symbols),
the sample of 16 $\ge$3.5-$\sigma$ SCUBA sources associated robustly
with MAMBO in the Lockman Hole and ELAIS\,N2 fields (blue symbols) and
the sample of MAMBO sources with SCUBA photometry presented by E03
(green symbols).  The open histogram shows the combined $\fdr$
distribution of all three samples, while the filled grey histogram
shows the distribution for the subset of radio-identified sources.  }
\label{figure:mambo-scuba-histograms}
\end{center}
\end{figure}

We begin by considering the sources which are robustly detected by
both MAMBO and SCUBA, i.e.\ sources which are in both of the deboosted source
catalogues {\em and} detected at $\ge$4\,$\sigma$ in the combined map.
In total, only five such sources were found (see
Table~\ref{table:mambo-scuba-source-list}) and their resulting $\fdr$
versus 850-$\mu$m flux density distribution is shown in
Fig.~\ref{figure:mambo-scuba-histograms}.

For comparison Fig.~\ref{figure:mambo-scuba-histograms}.
also shows the robust MAMBO/SCUBA sample
from E03, which are based on SCUBA photometry of MAMBO sources
extracted from surveys of the Lockman Hole, the NTT Deep Field and the
Abell\,2125 cluster field (Bertoldi et al.\ 2000; Dannerbauer et al.\ 2002;
2004). Values of $S_{\rm 850\mu m}$ and $S_{\rm 1200\mu m}$ in E03 did
not require deboosting as they were measured using `photometry mode'
(see Ivison et al.\ 1998b) at radio and/or mm interferometric
positions. We also compare with the MAMBO/SCUBA samples derived from
blank-field surveys of the ELAIS\,N2 and Lockman Hole fields by S02
and G04. For the five MAMBO/SCUBA associations found in the ELAIS\,N2
field, we derived $S_{\rm 1200\mu m}$ using the `deboosting curve'
given in G04, while a uniform deboosting factor of 15 per cent was
adopted for $S_{\rm 850\mu m}$ (S02). In both cases, the deboosting
factors were within the total flux calibration errors.  Using a
dual-survey extraction technique, Ivison et al.\ (2005 -- hereafter
I05) identified eight MAMBO/SCUBA associations in the Lockman Hole --
in addition to the eight already presented by G04 -- and we have
included those in our comparison here. For all 16 associations in the
Lockman Hole we have adopted $S_{\rm 850\mu m}$ (deboosted) from the
SHADES survey (Coppin et al.\ 2006) which has a larger amount of data
on the Lockman Hole than the UK 8-mJy Survey (S02). Deboosted $S_{\rm
1200\mu m}$ came from G04.
 
Although we have only identified five $\ge$4-$\sigma$ MAMBO/SCUBA
associations in GOODS-N, and therefore are dealing with small number
statistics, the sources are seen to span a similar range in $\fdr$ and
$S_{\rm 850\mu m}$ as existing samples. Most sources in all three
samples exhibit a correlation between $\fdr$ and $S_{\rm 850\mu m}$.
There are, however, a few outliers, although the uncertainties on some
of these are substantial.

G04 (contrary to E03) did not find a significant fraction of
sources with low $\fdr$ values, but there is no discrepancy between
E03 and the G04+I05 sample compiled here.  This is due to the
deboosted $S_{\rm 850\mu m}$ values and the wider and deeper coverage
at 850\,$\mu$m.  As a result of the good agreement, we have combined
all three samples in Fig.~\ref{figure:mambo-scuba-histograms} in order
to improve the statistics. The combined flux ratio distribution, shown
as the empty histogram, is seen to be asymmetric around its peak at
$\fdr \sim 2.5$, with 12 out of 48 sources (25 per cent) with
$\fdr \ge 3$ and a shoulder of low $\fdr$ values, with 18/48 (38
per cent) of the sources lying at $\fdr\le 2$.

%
%

%
%
\begin{figure}[h]
\begin{center}
\includegraphics[width=1.0\hsize,angle=0]{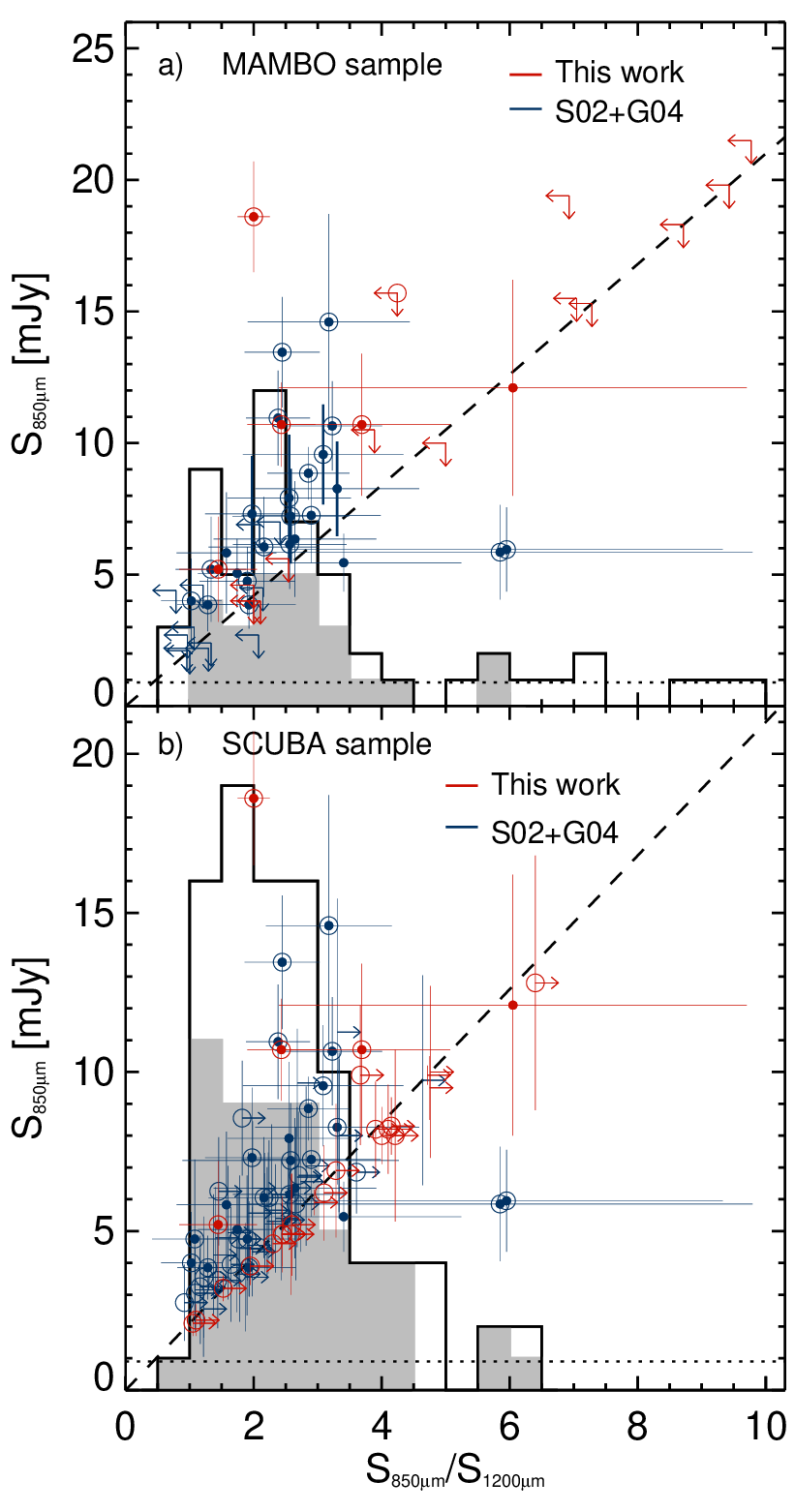}
\caption[]{{\bf a)} $S_{\rm 850\mu m}$ versus $\fdr$ for the entire
$\ge$4-$\sigma$ MAMBO sample {\it and} MAMBO sources detected at
$\ge$4\,$\sigma$ in the combined map (red symbols). For sources with
no robust SCUBA association we used the 3-$\sigma$ flux density limit
at the MAMBO position in the SCUBA map. The sample of MAMBO sources
obtained from the Lockman Hole and ELAIS\,N2 fields (S02+G04) with SCUBA
associations or upper limits on $S_{\rm 850\mu m}$ are also shown
(blue symbols).  The flux ratio distribution obtained by combining the
two samples is given by the open histogram, while the distribution of
radio-identified sources is shown as the grey histogram.  {\bf b)} The
opposite experiment from a): $S_{\rm 850\mu m}$ versus $\fdr$ for the
sample of $\ge$4-$\sigma$ SCUBA sources in GOODS-N {\it and} SCUBA
sources detected at $\ge$4\,$\sigma$ in the combined map (red
symbols).  SCUBA sources from S02+G04 with MAMBO counterparts or
$S_{\rm 1200\mu m}$ upper limits (3\,$\sigma$) are shown as blue
symbols. The open histogram shows the flux ratio distribution for the
combined GOODS-N and S02+G04 samples while the grey histogram shows
the distribution for only radio identified sources (see
\S \ref{subsection:mm-radio}).  In both a) and b), the dotted line
illustrates the 3-$\sigma$ detection limit in the deepest part of the
SCUBA map ($\sim$0.9\,mJy -- Pope et al.\ 2005), while the dashed line
shows the value of $S_{\rm 850\mu m}$ corresponding to the 3-$\sigma$
MAMBO detection limit ($\sim$2.1\,mJy) at a given flux density ratio.}
\label{figure:mambo-scuba-histograms-b}
\end{center}
\end{figure}

\subsection{Sources detected by MAMBO}
Next we expand our sample to include the $\ge$4-$\sigma$ MAMBO sources
(20 sources -- see Table \ref{table:mambo-source-list}) {\it plus}
MAMBO sources which were detected at $\ge$4\,$\sigma$ in the combined
map (GN\,1200.21, .27, .29, .33). The total tally is 18 sources within
the SCUBA region, with six sources outside (see bottom panel of
Fig.~\ref{figure:mambo-scuba-maps}). In cases where there was no
robust SCUBA association, we adopted the 3-$\sigma$ limit in the
850-$\mu$m map at the position of the MAMBO source. The resulting plot
of $S_{\rm 850\mu m}$ versus $\fdr$ is shown in
Fig.~\ref{figure:mambo-scuba-histograms-b}a. The uneven noise
properties of the SCUBA map are reflected in the large range of flux
density ratios exhibited by the MAMBO sample. A handful of MAMBO
sources in the shallowest parts of the SCUBA map have poorly
constrained flux ratios ($\ls$10). On the opposite side of the plot,
we find three sources with tight upper limits on $S_{\rm 850\mu m}$,
implying $\fdr\ls 2$. A number of such sources are also found when
repeating the analysis for the ELAIS\,N2 and Lockman Hole fields
(S02+G04). For these two fields the SCUBA and MAMBO maps were
well-matched in depth and, as a result, no sources with very high flux
density ratios are found.  Combining the S02+G04 results with those
from GOODS-N, we obtain the open histogram shown in
Fig.~\ref{figure:mambo-scuba-histograms-b}a.  Accounting for the upper
limits in the distribution, we find that 20--35 (37--65 per cent) of
the 54 sources in the combined distribution lie at $\fdr \ls 2$ while
9--18 (17-33 per cent) have $\fdr \gs 3$.

\subsection{Sources detected by SCUBA}
As an important check, we performed the mirror-image experiment,
deriving $\fdr$ for the SCUBA sample. We included all $\ge$4-$\sigma$
SCUBA sources (22 sources -- Pope et al.\ 2006) {\it and} sources
which were detected at $\ge$4\,$\sigma$ in the combined map
(GN\,850.39, .23, .28, .30 and .40), yielding a total of 27
sources. As expected, most of the sources lie along the dashed line in
Fig.~\ref{figure:mambo-scuba-histograms-b}b, which shows the flux
density ratio corresponding to the 3-$\sigma$ detection limit of the
MAMBO map ($\sim$2.1\,mJy) for a given 850-$\mu$m flux density.  The
flux ratio distribution obtained by combining the GOODS-N sample with
that of S02+G04 has a single dominant peak at $\fdr \sim 2-3$.  Taking
into account the lower limits in the combined distribution, we find
that 10--37 (11--39 per cent) of the 94 sources in the distribution
lie at $\fdr \le 2$, while 26--74 (28--79 per cent) lie at $\fdr \ge
3$.

A Kolmogorov-Smirnov (KS) test of the flux density ratio distributions
of the combined MAMBO and SCUBA samples in
Fig.~\ref{figure:mambo-scuba-histograms-b} yields a likelihood of
$\sim$46 per cent that identical distributions could differ by as much
as that observed\footnote{In performing the KS tests between the MAMBO
and SCUBA samples we have assumed that the upper and lower limits are
measured values.}, i.e.\ the distributions are indistinguishable
statistically. Nonetheless, we do note that 37--65 per cent of the
sources in Fig.~\ref{figure:mambo-scuba-histograms-b}a have $\fdr \le
2$ and 17--33 per cent have $\fdr \ge 3$, while the corresponding
percentages in Fig.~\ref{figure:mambo-scuba-histograms-b}b are 11--39
and 28--79 per cent, respectively, i.e.\ the fractions are almost
reversed between the two distributions.

Due to the uneven noise properties of the SCUBA map in particular, the
non-negligible errors on the (sub)mm flux densities, and the small
sample sizes, we are unable from
Fig.~\ref{figure:mambo-scuba-histograms-b} to make conclusive
statements about any difference between the $\fdr$ distributions of
MAMBO- and SCUBA-selected samples.
However, we will reinvestigate this
same question using two more fruitful approaches in
\S~\ref{section:stacking} and \ref{section:stat-mambo-scuba-maps}.  But
before doing so we will finish this section by examining
whether radio identifications are helpful here.

\subsection{850-/1200-$\mu$m flux ratio distribution of
radio-bright and radio-dim (sub)mm sources}
\label{subsection:mm-radio}

For the SCUBA samples in the Lockman Hole and ELAIS\,N2 regions we adopted the
radio-identified SMGs in Ivison et al.\ (2002, 2007). For the SCUBA
sample in GOODS-N, we adopted the radio identifications by Pope et
al.\ (2005, 2006) and Biggs \& Ivison (2006). To identify radio
counterparts to our MAMBO sample in GOODS-N, we used the Biggs \&
Ivison (2006) radio catalogue to search for sources within 6\,arcsec of each
MAMBO source. For every such association, the probability, $P$, of this
happening by chance was calculated using the method of Downes et al.\
(1986).  A radio identification was considered robust if $P <
0.05$. In Tables~\ref{table:mambo-source-list} and
\ref{table:mambo-scuba-source-list}, respectively, we list radio
identifications for the catalogues extracted from the MAMBO data
alone, and from the combined SCUBA/MAMBO map.

We find that only 9/30 (30 per cent) of the deboosted MAMBO sources
have robust radio identifications, significantly lower than the
typical identification fraction (65 per cent -- e.g.\ Ivison et al.\
2002) for bright SCUBA sources and radio data of similar depth ($\rm
\sigma_{1.4GHz}\sim 5\,\mu Jy\,beam^{-1}$). Considering only the
1200-$\mu$m sources detected at $\ge$4\,$\sigma$, the fraction
increases although to no more than 35 per cent. If we instead consider the
sample extracted at $\ge$4\,$\sigma$ from the combined map
(Table~\ref{table:mambo-scuba-source-list}), we find 17/33 radio
identifications (52 per cent), roughly consistent with that of bright
SCUBA sources.

From the combined samples in Fig.~\ref{figure:mambo-scuba-histograms}
we find that the fraction of MAMBO/SCUBA associations which also have radio
identifications is roughly constant ($\sim$60--100 per cent) across
the range of flux ratios spanned by the combined sample.  This is not
surprising since we have only considered sources with high-S/N SCUBA
and MAMBO detections. We therefore conclude that SMGs exhibit a broad
range of $\fdr$ values, including quite low ratios, independent of
whether or not they are identified at radio wavelengths.

Turning our attention to Fig.~\ref{figure:mambo-scuba-histograms-b}
and the flux ratio distributions of the MAMBO- and SCUBA-selected
samples, and in particular the fraction identified at radio
wavelengths as a function of flux density ratio (grey histograms), we
find that at low flux density ratios ($\fdr \ls 2$) the fraction of
MAMBO sources with radio identifications is 0--60 per cent, while the
corresponding fraction for the SCUBA-selected sample is 47--100 per
cent. Given that this fraction is expected to drop at $z\gs 3$, this
is the kind of scenario we would expect to see if 1200-$\mu$m
observations were selecting a significant population at those
redshifts. However, these are small number statistics; a formal KS test does not support the notion
that the two populations differ, giving a 24 per cent chance that the
two distributions are drawn from the same parent distribution.

\section{Stacking analysis}
\label{section:stacking}

As we saw in \S\ref{section:mambo-scuba-comparison}, only five
$\ge$4-$\sigma$ sources in the combined map of the GOODS-N region are
robustly detected at both 850- and 1200-$\mu$m. In this section we
shall examine whether the remaining MAMBO (SCUBA) sources with no
associations are detected at 850\,$\mu$m (1200\,$\mu$m) in a
statistical sense.

To this end, we have carried out a 1200-$\mu$m stacking analysis of
the 16 SCUBA sources in Table~\ref{table:mambo-scuba-source-list} with
no MAMBO associations, but which were extracted from the combined
SCUBA/MAMBO map at significance $\ge$4\,$\sigma$.  To avoid
contamination by bright MAMBO sources and their negative sidelobes we
produce a {\sc clean}ed version of the MAMBO map in which all the
$\ge$3.5-$\sigma$ deboosted sources (including their negative
sidelobes) were masked, as well as a region 11\,arcsec in radius
around each one. The 1200-$\mu$m signal and noise values were
extracted from the MAMBO map at the positions of the SCUBA sources,
and the (variance-weighted) mean flux density and noise values were
derived.  This yielded a total of 15 SCUBA sources with a stacked flux
density of $\langle S_{\rm 1200\mu m}\rangle = 1.0\pm 0.1$\,mJy (S/N =
10). The average $S_{\rm 850\mu m}$ for these 15 SCUBA sources was
$3.8\pm 0.2$\, mJy.  As a check, we repeated the stacking analysis for
the $\ge$3.5-$\sigma$ sample in
Table~\ref{table:mambo-scuba-source-list} and found good agreement
with the stacked flux density of the $\ge$4-$\sigma$ sample. As an
extra check, we performed the stacking analysis with 10 per cent of
the brightest and faintest of the 1200-$\mu$m flux density values
removed from the stack, again finding agreement.

In a similar manner, we derived the stacked 850-$\mu$m flux density of
the 11 MAMBO sources in Table~\ref{table:mambo-scuba-source-list} with
no SCUBA associations, but which were recovered at $\ge$4\,$\sigma$
from the combined map. To this end, we used the {\sc clean}ed version
of the SCUBA signal map (Pope et al.\ 2005) to ensure the result would
not be affected by the negative sidelobes around each source.  A
15-arcsec region around each $\ge$3.5-$\sigma$ source was also blanked
out. The final stack contained all 11 sources and yielded a stacked
flux of $\langle S_{\rm 850\mu m}\rangle = 1.7\pm 0.6$\,mJy (S/N =
2.8).  The average $S_{\rm 850\mu m}$ of the 11 MAMBO sources was
$2.4\pm 0.2$\,mJy.  We found that the stacked $S_{\rm 850\mu m}$ for
the $\ge$3.5-$\sigma$ sample was consistent with the $\ge$4-$\sigma$
sample and that removing the 10 per cent brightest and faintest sources
from the stack did not change the result. Our stacking analysis is
summarised in Table~\ref{table:stacking} where we have also listed the
range in $\fdr$ implied by our analysis.

To gauge the significance of these results, we performed Monte Carlo
simulations in which the stacked signal was determined in the same
manner as above, except with randomised source positions.  For the
randomisation, we only considered positions within the MAMBO/SCUBA
overlap region. Furthermore, since we wanted the simulations to
represent similar noise properties to those in the original stacking
analysis, each source was assigned a random position 20--80\,arcsec
from its original position. From 10,000 such runs -- each time with
different positions -- we found that the observed 1200-$\mu$m signal
occurs on $\le$0.001 per cent of occasions, while there is a 4 per
cent chance of obtaining the observed 850-$\mu$m stacked signal. This confirms
the high significance of the stacked 1200-$\mu$m signal and suggests
the 850-$\mu$m signal could occur by chance.

It is important to realise that these stacked fluxes are likely to be biased low, since the MAMBO and
SCUBA maps are unlikely to be perfectly aligned and since the
positional uncertainties of (sub)mm sources at low significance levels
are $\sim$4\,arcsec.  As a result, we repeated the entire stacking
analysis, this time using the peak flux within 10\,arcsec of each
source in the stack, rather than the flux density at the exact
position (see \S~\ref{section:mambo-scuba-comparison} for the reason
for choosing a 10-arcsec radius). This yielded a stacked flux of
$\langle S_{\rm 1200\mu m}\rangle = 1.8\pm 0.1$\,mJy (S/N = 18) for
the 1200-$\mu$m-blank SCUBA sources; for the 850-$\mu$m-blank MAMBO
sources we found $\langle S_{\rm 850\mu m}\rangle = 4.0\pm 0.7$\,mJy
(S/N = 5.7).

%
%
%
\begin{figure*}
\begin{center}
\includegraphics[width=0.5\hsize,angle=0]{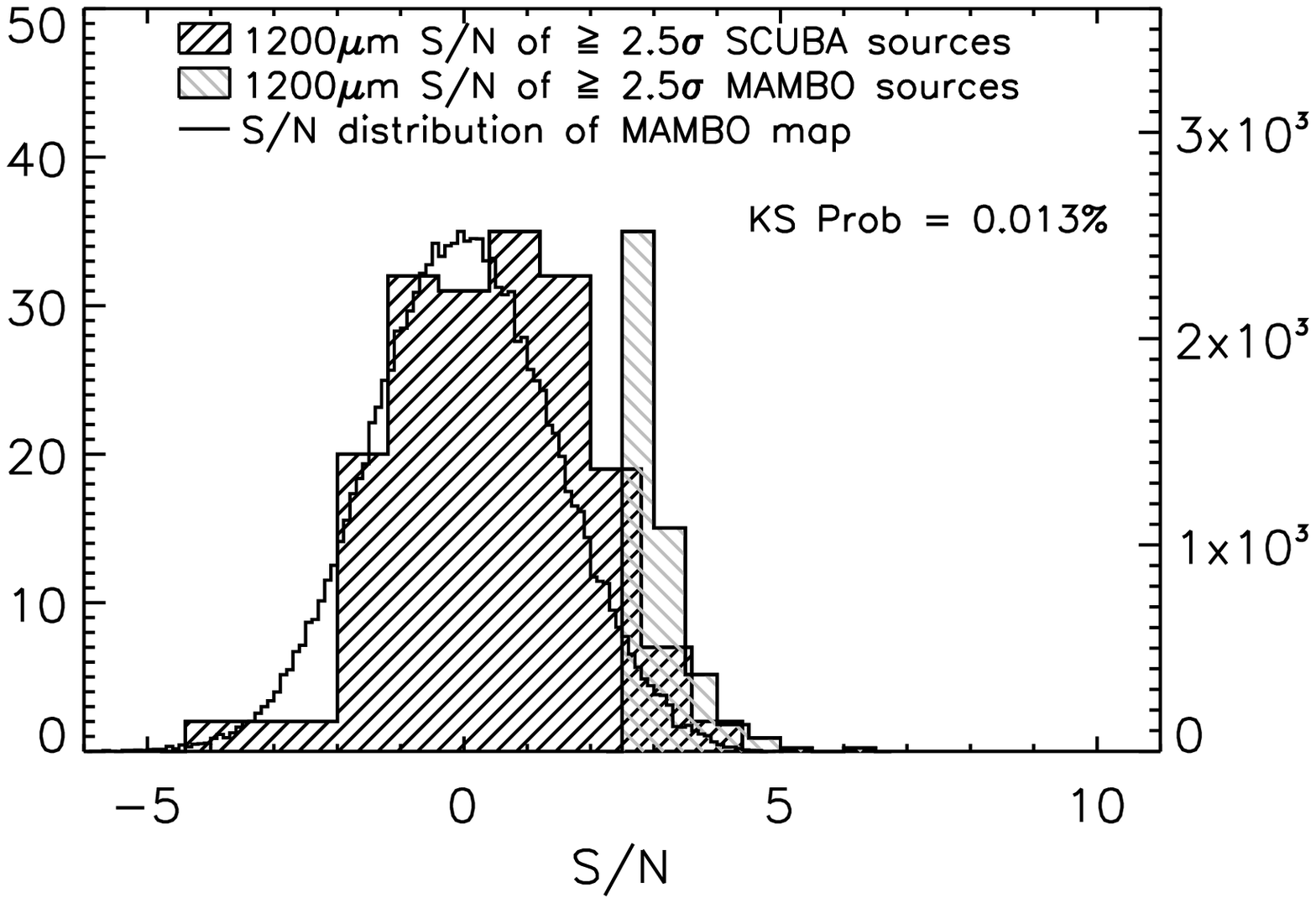}\includegraphics[width=0.5\hsize,angle=0]{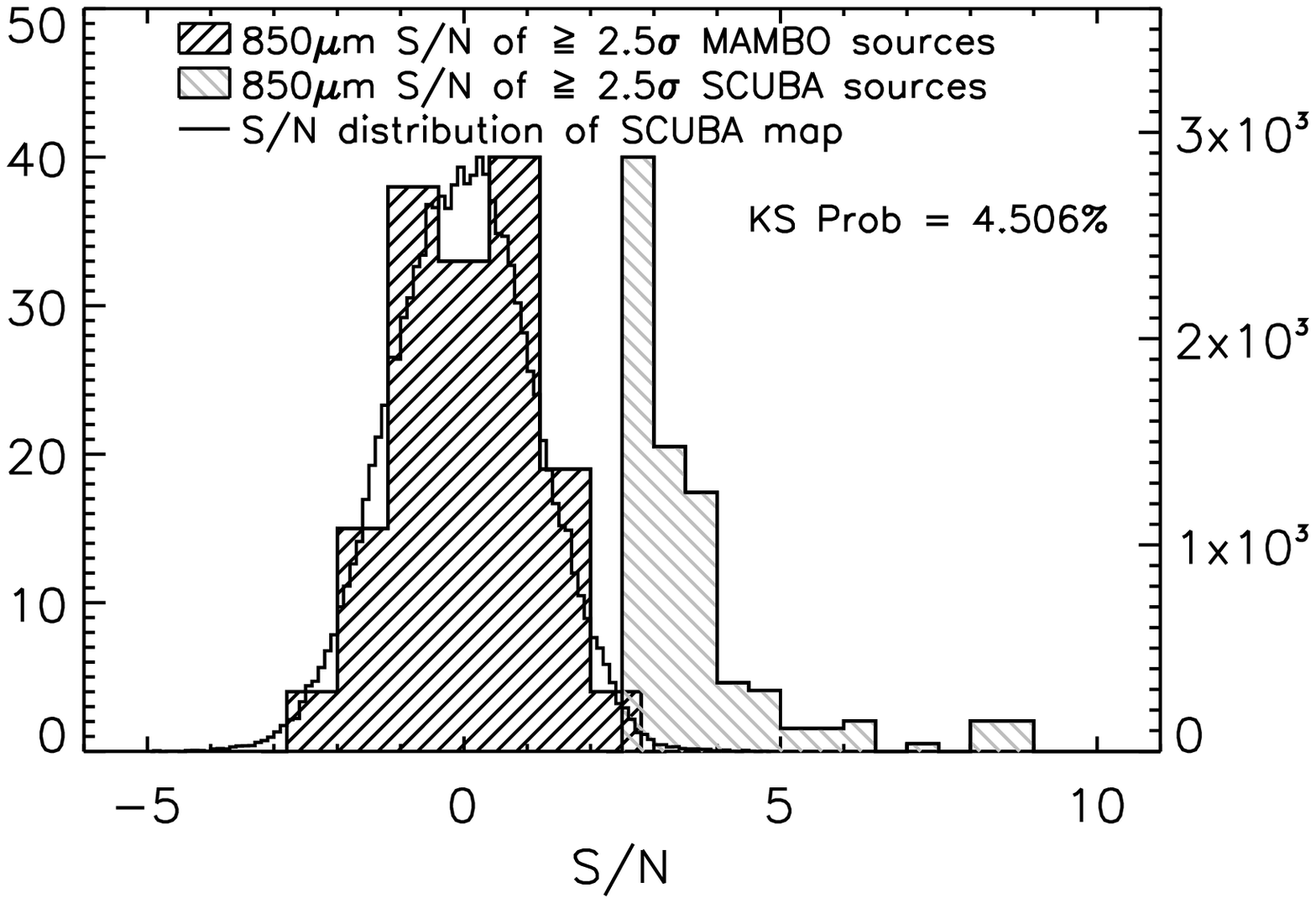}
\caption[]{{\bf Left:} Distribution of 1200-$\mu$m S/N values for the
$\ge$2.5-$\sigma$ SCUBA sources (hashed histogram and left $y$-axis)
compared to the 1200-$\mu$m S/N distribution of the {\sc clean}ed and
blanked MAMBO map (open histogram and right $y$-axis).  The former is
skewed towards positive values and a KS test yields a 0.013 per cent
likelihood that the two distributions are identical. Also shown for comparison is the
distribution of the 1200-$\mu$m S/N values of the $\ge$2.5-$\sigma$
MAMBO sources (grey hashed histogram), normalised to the histogram for
the $\ge$2.5-$\sigma$ SCUBA sources.  {\bf Right:} Distribution of the
850-$\mu$m S/N values of the $\ge$2.5-$\sigma$ MAMBO sources (hashed
histogram and left $y$-axis), compared to the 850-$\mu$m S/N
distribution of the {\sc clean}ed and blanked SCUBA map (open
histogram and right $y$-axis). The two distributions are drawn from the
same parent distribution with 4.5 per cent likelihood. For comparison,
the 1200-$\mu$m S/N distribution of the $\ge$2.5-$\sigma$ SCUBA
sources is also shown (grey hashed histogram).  }
\label{figure:snr-dist}
\end{center}
\end{figure*}

While this method yields an apparently significant 850-$\mu$m
detection of the 850-$\mu$m-blank MAMBO sources, selecting the peak
within 10\,arcsec of each source will bias the stacked 850-$\mu$m flux
density to high values. We consider the estimated signal from this
method to represent an upper limit, so the average 850-$\mu$m flux
density lies in the range $\langle S_{\rm 850\mu m}\rangle \simeq 1.7
- 4.0$\,mJy.  The SCUBA sample is detected significantly
($\ge$10\,$\sigma$) at 1200\,$\mu$m with both methods with $\langle
S_{\rm 1200\mu m}\rangle \simeq 1.0-1.8$\,mJy.

Stacked fluxes obtained by selecting peaks within a certain radius are
prone to flux boosting. We therefore repeated the above analysis, this
time deboosting the peaks using the Bayesian scheme described in \S~\ref{subsection:flux-boosting}. 
This yielded a stacked flux of
$\langle S_{\rm 1200\mu m}\rangle = 1.2\pm 0.1$\,mJy (S/N = 12) for
the 1200-$\mu$m-blank SCUBA sources, while for the 850-$\mu$m-blank MAMBO
sources we found $\langle S_{\rm 850\mu m}\rangle = 3.0\pm 0.7$\,mJy
(S/N = 4.3). The net
effect of deboosting is to lower the stacked flux estimates, narrowing
the range in $\langle S_{\rm 850\mu m}\rangle/\langle S_{\rm 1200\mu
m}\rangle$ for both the MAMBO and SCUBA samples (see Table \ref{table:stacking}).

%
%
\begin{table*}
\scriptsize
\caption{Summary of stacking results for the 1200-$\mu$m-blank SCUBA
sources and the 850-$\mu$m-blank MAMBO sources. Both the signals
stacked using the flux densities at the (sub)mm centroids (biased low)
and the peak flux densities within a 10-arcsec aperture (biased high)
are listed. The last column lists the stacked signals obtained when
deboosting the peak fluxes. The percentages in parentheses are the
likelihoods of obtaining the stacked fluxes by chance.}
\vspace{0.5cm}
\begin{center}
\begin{tabular}{llll}
\hline
\hline
\multicolumn{3}{c}{1200-$\mu$m-blank SCUBA sources (N=15)} \\ 
\hline
                                                                      &  At centroid                           & Peak (measured)                               &  Peak (deboosted)\\
\hline
$\langle S_{\rm 1200\mu m}\rangle$ /mJy                              &  $1.0\pm 0.1$ ($<10^{-3}$)  & $1.8\pm 0.1$ ($<10^{-3}$)  & $1.2\pm 0.1$ ($<10^{-3}$)\\
$\langle S_{\rm 850\mu m}\rangle$  /mJy                              &  $3.8\pm 0.2$                    & $3.8\pm 0.2$                    & $3.8\pm 0.2$\\
$\langle S_{\rm 850\mu m}\rangle/\langle S_{\rm 1200\mu m}\rangle$        &  $3.8\pm 0.4$                    & $2.1\pm 0.2$                    & $3.2\pm 0.3$\\
\hline
\hline
\multicolumn{3}{c}{850-$\mu$m-blank MAMBO sources (N=11)} \\ 
\hline
                                                                      &  At centroid                & Peak (measured)                        & Peak (deboosted)\\
\hline
$\langle S_{\rm 1200\mu m}\rangle$ /mJy                              &  $2.4\pm 0.2$        & $2.4\pm 0.2$            & $2.4\pm 0.2$      \\
$\langle S_{\rm 850\mu m}\rangle$  /mJy                              &  $1.7\pm 0.6$ (4)  & $4.0\pm 0.7$ (3)     & $3.0\pm 0.7$ (4)     \\
$\langle S_{\rm 850\mu m}\rangle/\langle S_{\rm 1200\mu m}\rangle$        &  $0.7\pm 0.3$        & $1.7\pm 0.3$            & $1.3\pm 0.3$      \\
\hline
\label{table:stacking}
\end{tabular}
\end{center}
\end{table*}

In addition to the stacking analysis, we compared the distribution of
850-$\mu$m S/N values at the positions of the 850-$\mu$m-blank MAMBO
sources with that of the overall S/N distribution of the ({\sc
clean}ed and blanked) SCUBA map, and similarly compared the
1200-$\mu$m S/N distribution of the 1200-$\mu$m-blank SCUBA sources
with the overall S/N distribution of the ({\sc clean}ed and blanked)
MAMBO map (only considering the region overlapping with the SCUBA
map). Comparing the full S/N distributions is likely to be a more robust
method than calculating the stacked flux densities, which only uses
the first moment. We performed this analysis for the 1200-$\mu$m-blank SCUBA
sources and 850-$\mu$m-blank MAMBO sources which had been extracted at
$\ge$2.5\,$\sigma$ from the SCUBA and MAMBO maps, respectively.  We
stress that by `1200-$\mu$m- (850-$\mu$m-)blank', we mean sources
which did not have a $\ge$2.5-$\sigma$ MAMBO (SCUBA) association
within 10\,arcsec of its position in the combined map.  The resulting
distributions are shown in Fig.~\ref{figure:snr-dist}. The 1200-$\mu$m
S/N distribution of the 1200-$\mu$m-blank SCUBA sources is clearly
skewed towards positive values relative to the map. A KS test gives a
0.013 per cent likelihood that the two distributions are
identical. For the 850-$\mu$m-blank MAMBO sources, the 850-$\mu$m S/N
distribution is not strongly inconsistent with that of the overall SCUBA map (KS test
probability, 4.5 per cent).

The statistical detection of the 1200-$\mu$m-blank SCUBA sources is
consistent with these sources lying just below the detection threshold
at 1200-$\mu$m, but being otherwise identical to the detected sources.
The non-detection of the 850-$\mu$m-blank MAMBO sources -- except when
non-deboosted peak values are used in the stack -- suggests that these
sources lie {\em well} below the detection threshold, being much fainter
than typical SMGs.  These results are also reflected in the average
$\fdr$ values. For the latter, $\fdr \ls 1.7$, while the
1200-$\mu$m-blank SCUBA sources have $\fdr$ values consistent with
SMGs which have robust detections at both 850 and 1200\,$\mu$m.  Our
findings therefore support the notion that a significant fraction of
MAMBO sources selected from blank-field surveys are SDOs, with cooler
1200--850$\,\mu$m colours.

\section{Simulations of the association fractions}
\label{section:stat-mambo-scuba-maps}

Is the observed fraction of SCUBA sources with MAMBO associations (and
vice versa) consistent with what we would expect given the properties
of the maps? Since the noise varies significantly across the SCUBA
map, we must address this question by means of 
careful Monte Carlo simulations of the associations, taking into
account the varying depths of the SCUBA and MAMBO maps.

\begin{figure*}
\begin{center}
\includegraphics[width=1.0\hsize,angle=0]{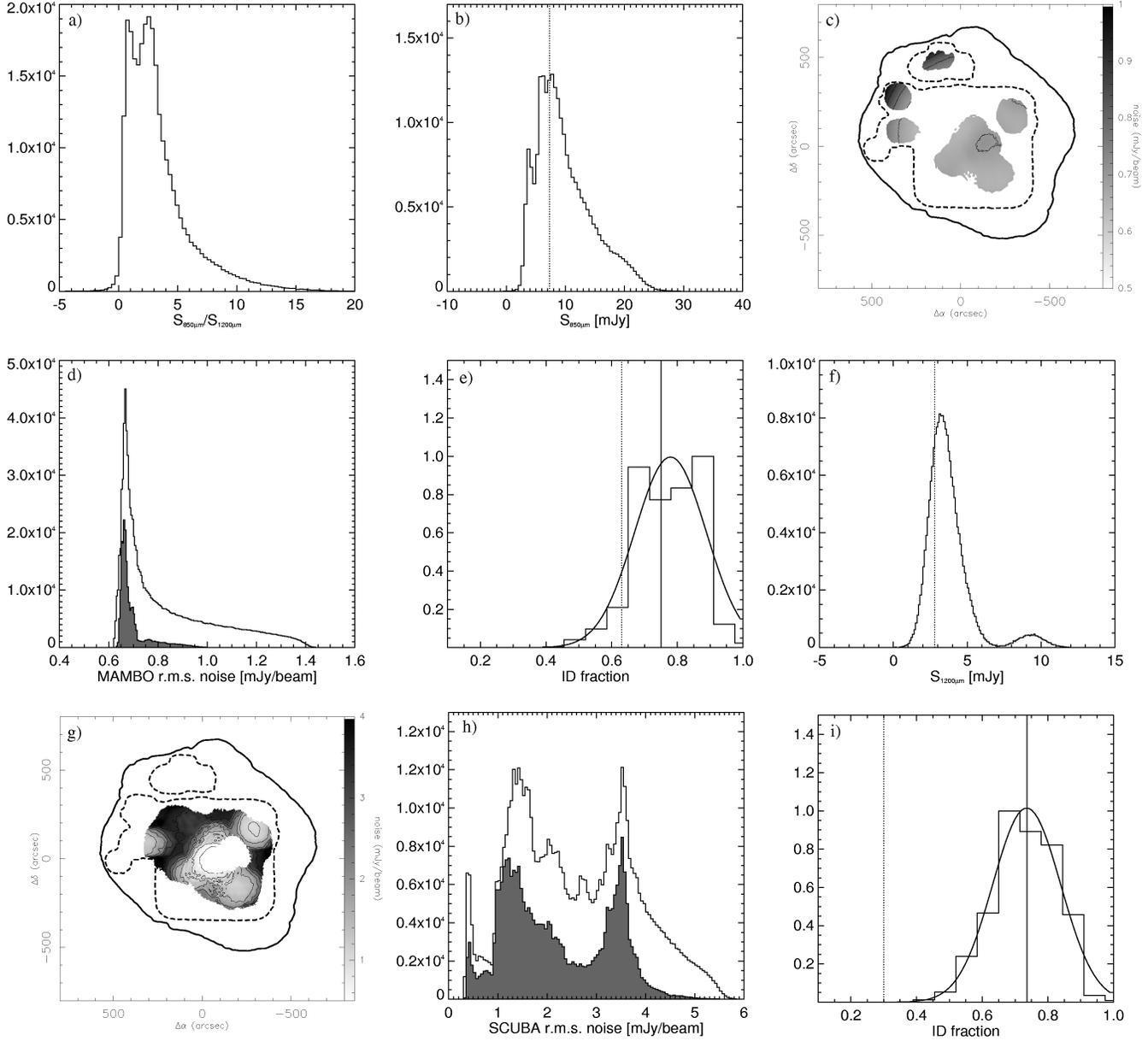}
\caption[]{Distributions from our Monte Carlo simulations, described
in \S\ref{section:stat-mambo-scuba-maps}.  {\bf a)} $\fdr$
distribution derived by resampling the flux density ratio distribution
of the 28 SCUBA/MAMBO associations found as described in \S~\ref{section:stat-mambo-scuba-maps}, taking into account the flux
density errors.  {\bf b)} The 850-$\mu$m flux density distribution
obtained by resampling the flux distribution of the $\ge$4-$\sigma$
SCUBA sources.  The vertical dotted line represents a flux density of
7.1\,mJy drawn from this distribution. {\bf c)} The grey-scale regions
show the parts of the MAMBO noise map which correspond to regions in
the SCUBA map where a flux density of 7.1\,mJy can be detected at
$\ge$4\,$\sigma$. This serves as an example of how we have taken the
inhomogeneous noise distribution of the maps into account.  {\bf d)}
The open histogram shows the noise distribution of the entire MAMBO
map, while the noise distribution of the regions selected in c) are shown
as a filled, grey histogram. {\bf e)} The resulting 1200-$\mu$m
association fraction distribution (normalised to unity) for the $\ge$4-$\sigma$ SCUBA
sample. The vertical solid line indicates the median of the
distribution (75 per cent), while the dotted line shows the actual
observed fraction (63 per cent). The solid curve represents a single
Gaussian fitted to the distribution.  {\bf f)} The flux density
distribution of the $\ge$4-$\sigma$ MAMBO sample. The vertical dotted
line represents a randomly chosen flux value of 2.8\,mJy.  {\bf g)}
The grey-scale regions show parts of the SCUBA noise map which
correspond to regions in the MAMBO map where a flux density of
2.8\,mJy can be detected at $\ge$4\,$\sigma$.  {\bf h)} The noise
distribution of the entire SCUBA map (open histogram), and the noise
distribution corresponding to the noise regions shown in g) (filled,
grey histogram).  {\bf i)} The 850-$\mu$m association fraction
distribution (normalised to unity) for the $\ge$4-$\sigma$ MAMBO sample. The vertical solid
line indicates the median of the distribution (74 per cent) and the
dotted line shows the actual observed fraction (30 per cent). }
\label{figure:mambo-scuba-stats}
\end{center}
\end{figure*}

To this end, we took the $\ge$2.5-$\sigma$ SCUBA and MAMBO catalogues
and searched for counterparts within 10\,arcsec of each other. In
total, 28 such SCUBA/MAMBO associations were found in this manner: 63
per cent of the $\ge$4-$\sigma$ SCUBA sources had 1200-$\mu$m
associations; 30 per cent of the $\ge$4-$\sigma$ MAMBO sources had
850-$\mu$m associations.  From the 28 SCUBA/MAMBO associations, we
constructed a $\fdr$ distribution (see
Fig.~\ref{figure:mambo-scuba-stats}a), taking into account the errors
(assuming Gaussianity). If SCUBA and MAMBO sources -- even at faint
flux density levels -- trace the same galaxy population, we would
expect the $\fdr$ distribution for the $\ge$2.5-$\sigma$ samples to be
representative for all SCUBA and MAMBO sources at all significance
levels. This is the null hypothesis we want to test.

First, we examined whether the observed fraction of SCUBA sources with
MAMBO associations is consistent with what we would expect. We ran a
Monte Carlo simulation in which we assigned a random flux density --
drawn from the resampled 850-$\mu$m flux density distributions of
$\ge$4-$\sigma$ SCUBA sources (Fig.~\ref{figure:mambo-scuba-stats}b)
-- to each member of the sample of $\ge$4-$\sigma$ SCUBA sources
detected with MAMBO at $\ge$2.5\,$\sigma$. Then a $\fdr$ value was
randomly selected from the $\fdr$ distribution, and the corresponding
$S_{\rm 1200\mu m}$ was calculated.

Next, we identified the regions of the SCUBA map where each 850-$\mu$m
flux density value could be detected at $\ge$4\,$\sigma$, given the
observed noise. Since the noise varies across the SCUBA map, so the
suitable regions vary for each flux value. We then drew random noise
values, $\sigma_{\rm 1200\mu m}$, from the noise distribution of the
parts of the MAMBO map which overlapped with the suitable SCUBA
regions.  In doing this, we have accounted for the varying noise
across the maps. An example of this is shown in
Fig.~\ref{figure:mambo-scuba-stats}c, where we show the regions of the
MAMBO noise map which overlap with the regions in the SCUBA map where
a flux density of 7.1\,mJy is detectable at $\ge$4\,$\sigma$.  The
noise distribution of this region is shown as the grey histogram in
Fig.~\ref{figure:mambo-scuba-stats}d.

Finally, the fraction of associations was calculated, based on how
many sources had $S_{\rm 1200\mu m}/\sigma_{\rm 1200\mu m} \ge 2.5$.
We ran this experiment 10,000 times, building up a distribution of
association fractions, where by `association fraction' we mean the
fraction of SCUBA sources with associations in the MAMBO image. From
the resulting distribution, shown in
Fig.~\ref{figure:mambo-scuba-stats}e, we see that that the observed
1200-$\mu$m association fraction (63 per cent) is fully consistent
(within 1\,$\sigma$) with the simulated distribution of association
fractions (median, 75 per cent). We ran two additional simulations,
similar to that described above, but using the $\ge$3.5- and
$\ge$5-$\sigma$ SCUBA samples. In both cases, we found that the
observed association fraction is consistent with the simulations. The
observed fraction of SCUBA sources recovered at 1200\,$\mu$m is
expected, then, given the noise properties of the maps.

%
%

Can the same be said for the 850-$\mu$m recovery fraction of MAMBO
sources?  To test this, we carried out the mirror-image experiment as
above, i.e.\ random 1200-$\mu$m fluxes were assigned to the sample of
$\ge$4-$\sigma$ MAMBO sources with $\ge$2.5-$\sigma$ SCUBA detections,
using the resampled $\ge$4-$\sigma$ 1200-$\mu$m flux distribution
(Fig.\ \ref{figure:mambo-scuba-stats}f). Using the same
850-/1200-$\mu$m distribution as above (i.e.\ derived from the
$\ge$2.5-$\sigma$ detections) the 850-$\mu$m flux densities were
determined for the sample. In a similar manner as before, the regions
of the MAMBO maps in which each 1200-$\mu$m flux density could be
detected at $\ge$4\,$\sigma$ were identified and 850-$\mu$m noise
values were randomly extracted from the the noise distributions of the
matching regions in the SCUBA noise map.  Finally, the fractions of
associations were calculated.  The resulting distribution of
850-$\mu$m association fractions for the $\ge$4-$\sigma$ MAMBO sources
is shown in Fig.~\ref{figure:mambo-scuba-stats}i and it is seen that
the observed 850-$\mu$m association fraction (30 per cent) of MAMBO
sources falls completely outside the simulated distribution of
association fractions, deviating from the mean (74 per cent) by
4\,$\sigma$.

The conclusion of this analysis is that, provided we are prepared to use the sample of lower significance
peaks, we can reject the null hypothesis for MAMBO sources. In other words
there is evidence for a population of 1200-$\mu$m sources which does
not share the distribution of flux density ratios with sources found
jointly at 850 and 1200\,$\mu$m. The non-trivial distribution shown in
Figs.~\ref{figure:mambo-scuba-stats}e and i underscores the importance
of performing these careful Monte Carlo tests in order to be
confident of this result.  In order to explain the low fraction of
associations, given the signal and noise properties of the two maps,
the 850-/1200-$\mu$m flux density distribution for the MAMBO
population would have to be skewed towards lower values. This argues
in favour of a significant fraction of them being SDOs, at higher
redshifts and/or with cooler dust.

\section{Discussion}
\label{section:discussion}

The spectroscopic redshift survey of bright, radio-identified SMGs by
Chapman et al.\ (2003, 2005) located sources out to $z\sim 3.6$, with
an interquartile range, $1.7\le z \le 2.8$. I05 argued that of the
bright SMG population ($S_{\rm 850\mu m} \ge 5$\,mJy), probably no
more than $\sim$10 per cent could be at $z > 3.5$, given that $\sim$80
per cent have radio counterparts and $\sim$10 per cent were likely to
be spurious. Pope et al.\ (2006) reached a similar conclusion,
estimating that $\ls$14 per cent of SMGs reside at $z\gs 4$. Modeling
the radio/mm/far-infrared colours of 120 SMGs from SHADES and adopting
priors for the redshift probability of their radio-undetected sources,
Aretxaga et al.\ (2007) argued that more than half of the bright SMG
population lies in the range $1.6\ls z \ls 3.4$, with little room for
a $z > 4$ population. Thus, the evidence for a {\it significant}
population of SMGs at very high redshifts appears slim, but has not
been ruled out completely.

If we consider sources {\it strictly selected at 850\,$\mu$m}, seven
have spectroscopically confirmed redshifts in the range $3 \ls z \ls
3.6$ (Ledlow et al.\ 2002; Chapman et al.\ 2003, 2005). A strong
candidate for a $z\gg 3$ SMG was reported by Knudsen, Kneib \& Egami
(2006) who inferred a likely spectroscopic redshift of $\simeq 4$ for
SMM\,J16359$+$66130. Apart from this source, however, only one other
convincing candidate for a $z > 4$ SMG has been presented in the
literature at the time of writing, namely GN\,850.10 (Wang et al.\
2004; Pope et al.\ 2005). This source was deemed to reside at $z\sim
4-6$, based on its near- and mid-infrared colours, which were
facilitated by its accurate location at 890-$\mu$m with the SMA (Wang
et al.\ 2007), at 1.1\,mm with the IRAM PdBI and in the radio
with the VLA (Biggs \& Ivison 2006; Dannerbauer et al.\ 2008).  Younger et al.\
(2008) presented SMA observations of LH\,850.2 -- one of the brightest
850-$\mu$m (and 1200-$\mu$m) sources in the Lockman Hole (G04; Coppin
et al.\ 2006) -- and derived an optical/infrared photometric redshift
of $z\simeq 3.3$, at the high-end of (but within) the Chapman et al.\
redshift distribution.  Aretxaga et al.\ (2007) presented a handful of
SMGs from SHADES with photometric redshifts $z\gs 4$, but with very
large individual uncertainties.

%
%
\begin{figure}
\begin{center}
\includegraphics[width=1.0\hsize,angle=0]{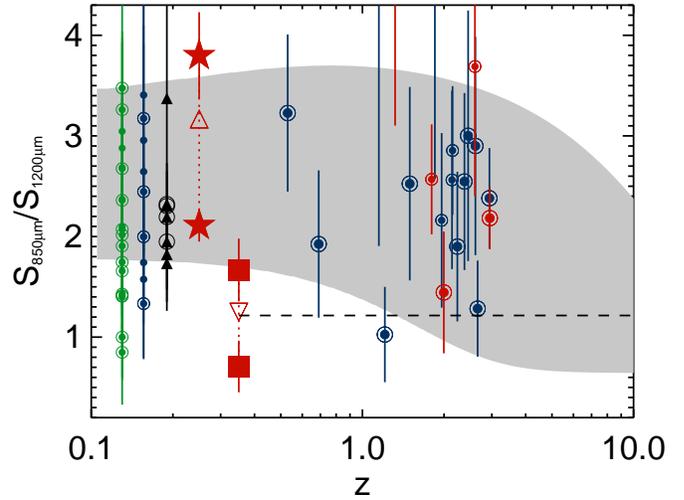}
\caption[]{Green, blue and red filled circles represent $\fdr$ for
sources identified robustly at 850 and 1200\,$\mu$m by E03, S02+G04
and in this work, respectively.  The filled triangles are the seven
sources from Younger et al.\ (2007), where we have converted the
measured 890- and 1100-$\mu$m flux densities to 850- and 1200-$\mu$m
flux densities assuming an optically thin grey-body with
$\beta=1.5$. Sources which have robust radio counterparts have been
circled.  Where available, we have placed sources at their
spectroscopic (Chapman et al.\ 2005) or photometric redshifts
(Aretxaga et al.\ 2007; Pope et al.\ 2006; Wall, Pope \& Scott 2008),
otherwise the sources have been placed at fixed redshifts below
$z=0.3$ for clarity. Sources with spectroscopic redshifts are shown as
large symbols while sources with only a photometric redshift -- or
lacking a redshift estimate entirely -- are shown as small symbols.
The grey shaded region corresponds to the range of $\fdr$ spanned by
grey-bodies with $\beta = 1.0$ and $T_{\rm d} = 10$\,{\sc k} (lower
limit) and $\beta = 2.0$ and $T_{\rm d} = 70$\,{\sc k} (upper
limit). Large red squares and stars represent the range in $\fdr$
derived from the (centroid and peak) stacking analysis for the
850-$\mu$m-blank MAMBO sample and the 1200-$\mu$m-blank SCUBA sample,
respectively.  The open upward- and downward-pointing triangles
represent the lower and upper limits on the flux density ratio,
respectively, obtained when using deboosted peak values in the
stacking analysis (see \S~\ref{section:stacking}).  }
\label{figure:S850S1200S14-z}
\end{center}
\end{figure}

Turning our attention to sources {\it strictly selected at mm
wavelengths}, we find a larger number of possible $z > 4$ candidates.
Using IRAM PdBI to accurately locate their positions, Dannerbauer et
al.\ (2002, 2004) presented three examples of bright ($S_{\rm 1200\mu
m} \gs 3.5$\,mJy) MAMBO sources which, based on their radio, (sub)mm
and near-infrared properties, were argued to lie at $z\gs 4$. The
non-detection of these sources with {\it Spitzer}/IRS spectroscopy
(Valiante et al.\ 2007) is consistent with this claim.  Younger et
al.\ (2007) presented 890-$\mu$m interferometric SMA observations of
seven very bright ($S_{\rm 1100\mu m} \ge 7.6$\,mJy) AzTEC sources
selected at 1100\,$\mu$m and argued that the five radio-dim ($S_{\rm
1.4GHz}\le 41\,\mu$Jy) sources in their sample were likely to reside
at higher redshifts than the radio-identified sources due to their
systematically higher submm-to-radio flux ratios, lower IRAC
3.6--8.0-$\mu$m flux densities and non-detections at 24\,$\mu$m.

Common to all of these $z > 4$ candidates, whether they are SCUBA-,
MAMBO- or AzTEC-selected sources, is that they are very bright (i.e.\
luminous) and have been detected at both submm and mm wavelengths.
The trend that the high-redshift candidates tend to be very bright was
first noted by Ivison et al.\ (2002), who from deep radio imaging of a
large sample of SMGs found evidence that very bright ($S_{\rm 850\mu
m}\gs 8$\,mJy) SCUBA sources tend to have higher submm-to-radio flux
density ratios than less luminous SMGs, indicating that they are at
higher redshifts (Carilli \& Yun 1999) and that strong luminosity
evolution may have taken place (see also Pope et al.\ 2006).

From the above summary (and following \S~\ref{section:stat-mambo-scuba-maps}),
one wonders whether submm- and mm-selected
sources are drawn from identical populations? Does the preponderance
of $z > 4$ candidates amongst the mm-selected sources reflect a real
difference between the two populations, or is this merely a
coincidence? The multi-wavelength properties of mm-selected sources
have been studied far less thoroughly than SMGs, because they were
discovered more recently and are often thought to be mere mm-wave analogues to SMGs. In
particular, no systematic spectroscopic surveys have been undertaken
of mm-selected sources and their redshift distribution is therefore not known.

A related question is: why are all of the $z > 4$ candidates amongst the brightest (sub)mm
sources known? Is this merely a selection effect, reflecting the fact that we
can currently detect only the brightest sources with (sub)mm and radio
interferometers, or does it reflect a genuine evolutionary effect
(Wall, Pope \& Scott 2007)?

In \S~\ref{section:stacking} we showed that while the 1200-$\mu$m-blank
SCUBA sources are statistically detected at 1200\,$\mu$m (S/N $\sim$
10), the 850-$\mu$m-blank MAMBO sources are not robustly detected at
850\,$\mu$m (S/N $\sim$ 2.8). In Fig.~\ref{figure:S850S1200S14-z} we
plot $\fdr$ for these two samples, using the stacking results listed
in Table~\ref{table:stacking}. The ranges of possible values for the
average $\fdr$ ratio for the two samples only barely overlaps, even when
accounting for the high and low biases introduced when using the
centroid and peak values, respectively (see \S
\ref{section:stacking}).  If we deboost the peak fluxes which go into the
stack, the allowed ranges do not overlap at all. On average,
therefore, the 850-$\mu$m-blank MAMBO sample has significantly lower
$\fdr$ ratios than the 1200-$\mu$m-blank SCUBA sample.

The strongest support for this result was given in \S~\ref{section:stat-mambo-scuba-maps}, based on Monte Carlo simulations
of the association fractions in the two maps.  Since we are working in
the low-S/N regime, the fraction of sources recovered in two different
maps will never be unity, so it is crucial to simulate the
experiments. Careful simulations are required because of the
inhomogeneous noise in the maps. The main outcome of our Monte Carlo
simulations is that while the fraction of SCUBA sources detected with
MAMBO is consistent with the $\fdr$ distribution of sources found in both maps,
there are more MAMBO sources {\it
not} detected by SCUBA than would be expected. Since the simulations
account for the noise properties of the two maps, we were able to
estimate the fraction of MAMBO sources which are undetected by SCUBA
(compared with the expectation if the populations were the same, but in the
presence of the observed noise properties) and found
this to be $\sim$40 per cent.

Thus, we have presented strong evidence which suggests that -- while
there is substantial overlap between the 850-$\mu$m- and 1200-$\mu$m-
selected sources in GOODS-N -- the two populations {\it are not}
identical. A significant fraction of the 1200-$\mu$m-selected sources
are unaccounted for by the 850-$\mu$m data. Due to the statistical
nature of our analysis and the careful way in which we have taken into
account the uneven depths and noise properties of the two maps of the
GOODS-N region, we are confident that this is a general result that reflects
the existence of two distinct, albeit overlapping, populations.

We still need to establish whether SDOs are genuinely at $z \gs 4$ or whether
they have cooler dust temperatures at redshifts typical of
radio-identified SMGs. Due to the degeneracy between dust temperature
and redshift (Blain 1999), and without direct spectroscopic redshifts,
it is not trivial to determine which of the two explanations is
appropriate.

We note from Fig.~\ref{figure:S850S1200S14-z} that if the typical SEDs
of SDOs are given by $T_{\rm d} = 10$\,{\sc k} and $\beta=1.0$, which
is believed to be extreme not only compared to starbursts at the
present day but also to SMGs (Kov\'{a}cs et al.\ 2006), then their
most likely average flux ratio ($\fdr\sim 1.2$) would indicate an
average redshift of $z\sim 1.3$. One such source is seen in
Fig.~\ref{figure:S850S1200S14-z}, namely Lock850.27/LH\,1200.7 (G04;
I05) which has a flux density ratio of $\sim 1$ and lies at
$z=1.21$. Given the general lack of spectroscopic redshifts for SMG
samples, further examples of such sources may exist, making them
viable candidates for SDOs. Adopting $T_{\rm d} = 20$\,{\sc k} and
$\beta = 1.5$, close to the coolest dust temperature measured for SMGs
so far (Kov\'{a}cs et al.\ 2006), results in an average redshift of
$z\sim 4.6$. If we assume that SDOs have SED properties similar to
those of radio-identified SMGs, i.e.\ $T_{\rm d} = 35$\,{\sc k} and
$\beta = 1.5$ (Kov\'{a}cs et al.\ 2006), we find their average
redshift to be $\sim 9$, in the regime where the $k$-correction
starts to become positive (Blain \& Longair 1993).

In the first of the three scenarios outlined above, we note that the
angular sizes of cool ($T_{\rm d} \ls 25$\,{\sc k}) SMGs at $z< 2$
should scale roughly as $\theta \sim (S_{\rm 850\mu m}/\mbox{mJy})^{1/2}$
(Kaviani et al.\ 2003) which would imply that SDOs have typical sizes
of $\theta \sim 1.3$\,arcsec (for $S_{\rm 850\mu m}=1.7$\,mJy -- see
Table~\ref{table:stacking}), corresponding to a physical diameter of
11\,kpc at $z=1.5$.  Thus if SDOs reside at $z\ls 2$ they would have
larger physical sizes than those observed for SMGs ($\sim$4\,kpc --
Tacconi et al.\ 2006; Biggs \& Ivison 2008), but similar angular sizes
($\sim$1\,arcsec). If SDOs predominantly lie at the same redshifts as
SMGs ($z\sim 2.5$) but have cooler dust temperatures ($T_{\rm d}\ls
25$\,{\sc k}), their angular sizes would be larger by a factor of two
(Kaviani et al.\ 2003).  Thus, if SDOs are in fact very cool systems
with a redshift distribution similar to SMGs (or peaking at lower redshift), we would
expect them to be quite extended in high-resolution radio and mm
images.

If some SDOs, on the other hand, lie in the range $z \sim 4-10$
and have typical dust temperatures in the range $T_{\rm d} \sim
20-40\,${\sc k}, the expected angular and physical sizes are $\theta
\sim 1.5-2.3$\,arcsec and 6.4--18.0\,kpc, respectively, inconsistent
with the typical sizes measured for SMGs.  Thus, if the Kaviani et
al.\ (2003) results are applicable, it would suggest that {\it SDOs are
very extended systems over a broad range of plausible dust
temperatures and/or redshifts.}  We note that high-resolution 1.3-mm
interferometry of SDOs should be able to test this and may provide a
useful way of discriminating between typical SMGs -- generally compact
at mm wavelengths ($\ls$1\,arcsec -- Tacconi et al.\ 2006; Younger et
al.\ 2007) and SDOs, which our analysis here suggest may be larger.

Observational evidence for such extended systems was recently
uncovered by Daddi et al.\ (2008), who demonstrated the existence of
gas-rich disks at $z=1.5$ with physical sizes 2--3 times those of
SMGs, as implied by CO observations as well as rest-frame UV light.
Some of these $BzK$ galaxies -- so called for the way they are
selected (Daddi et al.\ 2004) -- have been detected at very faint flux
levels at 1200-$\mu$m ($\sim$1.5\,mJy -- Dannerbauer et al.\ 2005) and
have similar star-formation efficiencies to those of local spirals,
i.e.\ an order of magnitude smaller than SMGs. They could be
candidates for the cold, extended `cirrus' dust models proposed by
Efstathiou \& Rowan-Robinson (2003). Certainly, these properties seem
to fit with those of SDOs, opening up the interesting possibility of
an overlap between the two populations, with SDOs making up 
some fraction of the near-infrared-selected population at
$z\sim 1.5-2.4$.

However, it seems clear from a growing number of in-depth studies, that some
of the brightest mm- (and submm-)selected sources are almost certainly
at $z > 4$ (Dannerbauer et al.\ 2002; Wang et al.\ 2004; Younger et
al.\ 2007). The same may be true for many of the SDOs, although we
stress that one cannot rule out a scenario in which SDOs are
predominantly cool, low-redshift sources.

Adopting the high-redshift scenario ($z\simeq 5$) for SDOs, and
assuming $T_{\rm d}=20\,${\sc k}, $\beta =1.5$ and $S_{\rm 850\mu m} =
1.7-4.0$\,mJy (Table~\ref{table:stacking}), we find far-infrared
luminosities and dust masses of $L_{\rm FIR} = (1-2)\times
10^{12}\,\Lsolar$ and $M_{\rm d} = (3-8)\times 10^{9}\,\Msolar$,
respectively. In this scenario, SDOs are luminous and massive
systems, despite being relatively fainter at submm wavelengths. The presence of
such systems at $z\simeq 5$ (corresponding to $\sim$1\,Gyr after the
Big Bang) would have implications for our understanding of galaxy
formation and evolution. Due to their large masses and high redshifts,
SDOs and SMGs at $z > 4$ would be potential candidates for galaxies
caught in the act of collapsing early in the Universe's history
(Eggen, Lynden-Bell \& Sandage 1962).

\section{Summary}
\label{section:summary}

We have undertaken the first deep and uniform 1200-$\mu$m survey of
the GOODS-N region, identifying 30 sources (after flux deboosting) in
our image. By combining our 1200-$\mu$m MAMBO map with the existing
850-$\mu$m SCUBA data, we have extracted a robust sample of 33 sources
(sub)mm sources detected at a combined S/N$\,{\ge}\,$4.

The principal question we wanted to address in this paper was whether
the 850- and 1200-$\mu$m selected sources constitute identical, or
merely overlapping, populations. We performed three independent
statistical analyses of the 850- and 1200-$\mu$m maps and their
corresponding source catalogues.

The first -- a simple comparison of the 850-/1200-$\mu$m flux density
ratio distributions for the 850-$\mu$m- and 1200-$\mu$m-selected
samples, suggests that a larger fraction of MAMBO sources have low
values of $\fdr$ compared to the SCUBA sources.  This method suffers,
however, from small sample sizes and the effects of maps with uneven noise properties,
and so we were unable to see a significant difference between the two
populations.

However, using the other two methods -- both of which take into
account the uneven noise properties -- we do find evidence to suggest
that the source populations selected by MAMBO and SCUBA are not
identical. The strongest evidence comes from our Monte Carlo analysis
of the fractions of SCUBA galaxies with MAMBO associations, and vice
versa, which showed that while the fraction of SCUBA sources with
MAMBO associations was consistent with the map properties and the
$\fdr$ distribution of sources robustly detected at both wavelengths,
the fraction of MAMBO sources with SCUBA associations was
significantly lower than expected. In fact, we found that about 40 per cent 
of the MAMBO sources were not recovered at 850\,$\mu$m, thus lending
strong evidence to the notion that the two populations are not identical.

We have argued that the average $\fdr$ of SDOs is significantly lower
than the bulk of the SMG population, suggesting that these sources are
either very cool or lie at higher redshifts.  It would require
extremely cold and unusual ($T_{\rm d} \simeq 10$\,{\sc k} and $\beta
\simeq 1$) far-infrared/mm SEDs to explain the presence of SDOs at low
redshifts ($z\ls 2$). Nonetheless, examples of such sources do
exist. We hint at a link between such cool, extended mm-selected
sources and the near-infrared-selected population of $BzK$ galaxies,
which are known to be gas-rich yet have star-formation rates an order
of magnitude lower than those of SMGs.

If we adopt more realistic SED parameters ($T_{\rm d} \simeq 20$\,{\sc
k} and $\beta \simeq 1.5$) for SDOs, it would imply an average
redshift of $z \gs 4$, i.e.\ beyond the range probed by spectroscopic
surveys so far. If this is the case, it is tempting to speculate that
SDOs could be potential candidates for galaxies collapsing very early
on in the Universe's history.

It should be stressed, however, that our analysis merely shows that SDOs are at
at generally higher redshifts (or are cooler) than SMGs.  Extensive comparison
with detailed models will be required in order to determine what range
of redshift distributions are consistent with the available data.  This was
beyond the scope of the present study.  This task will also be made much
easier with additional information on these sources, particularly spectroscopic
or optical/IR photometric redshifts.  However, this first requires finding
an identification of these SDOs at other wavelengths.

Deep interferometric observations at (sub)mm wavelengths sensitive
enough to probe the faint flux levels of SDOs, are likely to become one of the most
promising ways of furthering our understanding of SDOs. Such
observations would allow us to
identify near-/mid-infrared and optical host galaxies with which to
constrain their redshifts.  Due to the limited sensitivity of current
(sub)mm interferometers, however, such studies have been limited to
the brightest SMGs.  With the new generation of wide-band receivers,
and with ALMA on the horizon, studies of larger samples of SDOs are becoming feasible.
Furthermore, the wide, deep, multi-colour surveys planned with SCUBA-2
and {\sl Herschel\/} (and on a somewhat longer time-scale, with CCAT)
may provide useful far-IR/submm photometric redshifts for huge
samples of SDOs and SMGs (Hughes et al.\ 2002; Aretxaga et al.\ 2003).
Armed with these redshift constraints, heterodyne receivers with
extremely high bandwidth could be employed to search for redshifted CO
and/or C\,{\sc ii} line emission. For a complete census of obscured
star formation and AGN activity across cosmic time, we must determine
the range of SED properties for (sub)mm-selected sources, along with
their true redshift distribution. 
Our evidence that mm-selected galaxies may extend beyond $z=4$ suggests that
this may be an efficient way to select and study massive galaxy formation at
the earliest times.

\section*{Acknowledgments}

We are grateful to the IRAM staff at Pico Veleta, in particular to
Stephane Leon for his assistance with the pool observing. We thank
Helmut Dannerbauer and Alejo Mart\'inez Sansigre for discussions and
for useful comments on the paper.  We also thank the referee for a
useful, thorough and swift response.  AP and DS would like to thank
the Natural Sciences and Engineering Research Council of Canada. AP
also acknowledges support provided by NASA through the {\it Spitzer} Space
Telescope Fellowship Program, through a contract issued by the Jet
Propulsion Laboratory, California Institute of Technology under a
contract with NASA.

\end{document}